\documentclass[a4paper,11pt]{article}
\usepackage{jheppub} % for details on the use of the package, please see the JINST-author-manual
\usepackage{lineno}
\usepackage{xcolor}
\definecolor{dye}{rgb}{0.0, 0.2, 0.42}
\usepackage[normalem]{ulem}
\usepackage{booktabs}
\title{%\boldmath 
Geometry of non-Gaussianity in transient non-attractor inflation
}

 \author{Mohammad Hossein Namjoo and}
 \author{Bahar Nikbakht}
 \affiliation{ School of Astronomy, Institute for Research in Fundamental Sciences (IPM),\\Tehran, Iran, P.O. Box 19395-5531}

\emailAdd{mh.namjoo@ipm.ir}
\emailAdd{bahar.nikbakht@ipm.ir}

\abstract{
Inflationary models predicting abundant primordial black holes (PBHs) and large amplitude of scalar-induced gravitational waves (SIGWs) often rely on amplified fluctuations over limited scales, typically driven by phase transitions, particle production, or departures from slow-roll evolution. While the power spectrum of these models has been extensively studied, higher-order correlations are much less understood. Motivated by the complex physics involved and the fact that PBH and SIGW formation are both sensitive to non-linearities, we present a detailed study of the bispectrum as the leading non-linear effect in these scenarios. We refer to the scale- and shape-dependence of the bispectrum collectively as its {\it geometry}; and define a scale-dependent shape correlator to disentangle the two dependencies. Generally, we find that for the scales most affected by phase transitions and particle production (including the power spectrum peak), the bispectrum is strongest near the equilateral configuration, while non-attractor phases tend to produce correlations near the squeezed configuration. We further propose a simplified bispectrum estimator, resembling local-type non-Gaussianity but with scale-dependent amplitude, that captures the main features of the full bispectrum. As an implication of our results, we show that incorporating the bispectrum significantly broadens the range of scales with a substantial probability of large smoothed density contrasts compared to linear analysis. This suggests that non-linearities can alter not only PBH abundance and SIGW amplitude but also their mass and frequency spectra. In particular, and in contrast with the usual assumption, our results hint that the second-highest peak of the power spectrum may produce more PBHs than the highest peak.
}

\begin{document}
\maketitle
%\flushbottom

%%%%%%%%%%%%%%%%%%%%%%%%%%%%%%%%%%%%%%%%%%%%%%%%%%
%%%%%%%%%%%%%%%%%%%%%%%%%%%%%%%%%%%%%%%%%%%%%%%%%%
\section{Introduction}
\label{sec:Intro}
Single-field inflationary models with transient non-attractor phases have attracted much attention for various reasons. They are of particular interest for their potential to enable the abundant formation of primordial black holes (PBHs) \cite{Sasaki_2018, Biagetti_2018, Carr_2020, zsoy_2023, Escriv_2024}, and for the generation of scalar-induced gravitational waves (SIGWs) \cite{Domenech_2021,Pi:2020otn}. Such models typically cause a strong violation of scale-invariance and produce an enhanced power spectrum that sharply peaks at a specific scale, followed by rapid oscillations at smaller scales. These features are understood as being linked to the non-attractor evolution of the inflationary background and the transitions to and from the non-attractor phase. The peak scale is expected to determine the mass of PBHs and the frequency of SIGWs, thus has been the main focus in the majority of existing literature \cite{Byrnes:2018txb,Namjoo:2024ufv,Briaud:2025hra}. 

While most studies on such models are limited to the power spectrum, it is clear that going beyond the two-point correlation function is generally necessary, as non-linear (or even non-perturbative) effects are involved in both PBH and SIGW production \cite{Hooshangi:2021ubn,Adshead:2021hnm,Cai:2018dig}. The key quantity that captures the leading non-linear and non-Gaussian effects is the three-point function or bispectrum. Since scale-invariance is strongly violated in these models, the bispectrum is  expected to exhibit complex {\it geometry} (i.e., scale- and shape-dependence).\footnote{Intuitively, by the scale- and shape-dependence of bispectrum, we mean its dependence on the size and type of the triangle formed by the three momenta, respectively.}

 It has been common to consider a local-type bispectrum to quantify the leading non-Gaussianity near the peak of the power spectrum \cite{Cai_2022,Taoso:2021uvl,Escriva:2022pnz}. This may naively be justified by the expectation that the local-type non-Gaussianity emerges during non-attractor inflation \cite{Namjoo_2012}. However, the non-attractor evolution must be terminated soon enough, since otherwise the perturbations grow beyond the regime of perturbative control. That is why the non-attractor phases are assumed to be temporary, bracketed by slow-roll phases of inflation. Therefore, phase transitions are likely unavoidable in such scenarios, which account for deviations from scale-invariance. Generally speaking, since a narrow range of scales is strongly affected by a phase transition, it is reasonable to expect significant non-Gaussianity at the equilateral configuration within that range. It is unclear whether the combination of the assumption of local non-Gaussianity, along with the scale-dependence of the power spectrum, can fully explain the non-Gaussianities in such a complex situation. In fact, at least in scale-invariant scenarios, local non-Gaussianity is more prominent in the squeezed configuration (when one momentum is much smaller than the other two), rather than in the equilateral one. 
 
 Furthermore, as sharp phase transitions typically cause deviations from the Bunch-Davies (BD) vacuum, it may naively be expected that the bispectrum will also be sizable at the folded configuration (where the three momenta form an isosceles triangle with the base twice as long as the sides) \cite{Chen_2010}. This is because positive and negative frequency mixing can occur, unlike the pure positive frequency mode in the BD vacuum. 
 
 These considerations motivate a more detailed study of the bispectrum in transient non-attractor models, which is the main goal of this paper. While our method of analyzing the geometry of bispectrum is novel, a few papers exist where explicit calculation of the full bispectrum for transient non-attractor models is performed, which we comment on as we proceed \cite{Davies:2021loj,Taoso:2021uvl,Tasinato:2023ioq,Ragavendra:2023ret}.

In our previous work \cite{Namjoo:2024ufv}, we studied the scale-dependence of the squeezed-limit bispectrum of the curvature perturbation using a novel consistency relation advocated in Ref.~\cite{Namjoo:2023rhq}. In particular, we made generic statements about the sign, size and slope of the squeezed-limit bispectrum at the peak of the power spectrum. In this paper, we take one step further and study the full bispectrum, valid at all scales and configurations, in several models motivated by scenarios involving transient non-attractor phases. Our main focus is on the transient ultra-slow-roll (USR) inflation as a sufficiently generic, yet tractable, representative of such models \cite{Cai_2018}. The USR inflation happens when the inflaton field rolls on a constant potential, resulting in the production of significant non-Gaussianity and the violation of the slow-roll consistency relation \cite{Kinney_2005, Namjoo_2012,Passaglia_2019, _zsoy_2022,Cai_2022,Chen_2013}.

Consistent with our intuitive anticipation, the full bispectrum that we compute does show complex {\it geometry}. To disentangle and separately analyze the scale- and shape-dependence of a bispectrum, we define a scale-dependent shape correlate that measures the similarity between the full bispectrum and a set of well-motivated bispectrum shapes; each shape being significant at a specific configuration. Namely, we consider scale-invariant equilateral, local and folded shapes, and also introduce a new shape that is orthogonal (in the sense that we will make precise) to the local shape. We find that the full bispectrum typically shows significant overlap with the local and equilateral shapes at different scales, consistent with the intuition that the non-attractor phase generates local-type non-Gaussianity, while the phase transitions tend to generate equilateral non-Gaussianity. In particular, we find that at the scale where the power spectrum peaks (which is the perceived scale for the PBH and SIGW production), it is the equilateral configuration, rather than the squeezed one, that plays the major role. 

In contrast with the equilateral and squeezed configurations, we show that the folded non-Gaussianity is unlikely to be significant for the models under study. Despite the fact that both positive and negative frequency modes are involved, we understand this as a consequence of the transitions occurring in a short period of time, disallowing a significant mixing between the two modes. 

Since the bispectrum is generally quite complex, we also propose an estimator, dubbed here as {\it local-like bispectrum},  that is much simpler but still captures the main features of the bispectrum. The estimator resembles the bispectrum from local-type non-Gaussianity when the scale-dependence of the power spectrum (appearing in the local-type non-Gaussianity's expression) is taken into account and with the additional twist that the size of non-Gaussianity must be allowed to vary as a function of scale, calculable from the full bispectrum using a prescription we describe in this paper. Therefore, a purely local-type bispectrum cannot completely describe the three-point function for transient non-attractor models. 

The complex {\it geometry} of bispectrum can significantly impact PBHs and SIGWs. It is well-known that non-Gaussianities can significantly alter the abundance of PBHs and the energy density of SIGWs \cite{Byrnes_2012, Bullock_1997, Cai:2018dig,Pi_2025}. Our findings offers yet another effect: it can modify the range of scales responsible for their formation. This, in turn, implies that non-linear effects can reshape the mass and frequency spectrum of PBHs and SIGWs, respectively. As an approximate way to assess this, we calculate the probability of large local and smoothed density contrasts, considering the full bispectrum of the curvature perturbation. Notably, we find that including the full bispectrum not only substantially alters the likelihood of large fluctuations but also its scale-dependence. Additionally, we demonstrate that the scale exhibiting the largest probability can differ significantly from the peak scale of the power spectrum. In the transient USR model, this occurs because non-Gaussianities can become more prominent near the second-highest peak compared to the highest peak of the power spectrum. This leads the highest probability to be near the second-highest peak of the power spectrum. This challenges the previous assumption that the peak of the power spectrum and the peak of the probability of large fluctuations occur at the same scale; such an effect could substantially alter predictions of PBH masses and SIGW frequencies.\footnote{It is worth noting that the probability of PBH formation generally requires a non-perturbative treatment \cite{Hooshangi:2021ubn,Biagetti:2021eep}. To our knowledge, a self-consistent  non-perturbative method for analyzing  the models under consideration has not been developed; therefore, a complete picture remains elusive. However, as will be argued in Sec.~\ref{sec:beta}, accounting for the bispectrum's influence is expected to partially capture what a full analysis might reveal.}

The rest of this paper is organized as follows: In Sec.~\ref{sec:tools}, we introduce the tools and formalism required for our analysis, including the definition of the inner product and shape correlator. In Sec.~\ref{sec:examples}, we illustrate the use of these tools with two preliminary but insightful examples: a USR inflationary model with non-Bunch--Davies initial states, and a model involving a fast transition between two slow-roll phases. In Sec.~\ref{sec:transient USR}, we turn to the transient USR model, and examine the {\it geometry} of the full bispectrum. In Sec.~\ref{sec:estimator}, we discuss the estimator and the prescription for its application. In Sec.~\ref{sec:beta}, we prescribe a method to consider the impact of the full-shape bispectrum of the curvature perturbation to the statistics of the  smoothed matter overdensity and calculate the probability of large density contrast. We conclude in Sec.~\ref{sec:conclusion} and leave some technical details to Appendix.~\ref{sec:details}.

\section{Notation and tools}
\label{sec:tools}
In this section, we introduce the notation and the formalism used to quantify the degree of similarity between two shapes of primordial non-Gaussianity in non-scale-invariant scenarios. 

The main quantity of interest is the curvature perturbation on comoving slices, which we denote by $\mathcal{R}$. The unequal-time two-point function $\mathcal{R}$ can generally be expressed by
\begin{equation}
    \langle \mathcal{R}_{\mathbf{k}_1}(\tau_1) \mathcal{R}_{\mathbf{k}_2} (\tau_2) \rangle = \left( 2 \pi \right) ^3 \delta^3(\mathbf{k}_1 + \mathbf{k}_2) P_{k_1}(\tau_1, \tau_2)\, ,
\end{equation}
where $P_{k} (\tau_1 , \tau_2)$ denotes the unequal-time power spectrum, which in general is complex, except in the equal-time limit, i.e. $\tau_1 = \tau_2$. The observable two-point function is the equal-time limit of $P_{k}(\tau_1 , \tau_2)$ evaluated at the end of inflation, $P_{k} \equiv \lim_{\tau \to 0} P_k(\tau , \tau)$.
We will also use the power spectrum on logarithmic scales, defined as
\begin{equation}
    \mathcal{P}_k= \frac{k^3}{2 \pi^2} P_k.
\end{equation}

The bispectrum of the curvature perturbation $B_\mathcal{R}$ is related to its three-point correlation function, dictated by the spacetime symmetries, via
\begin{equation}
\label{eq:bispectrum}
\left\langle\mathcal{R}_{\mathbf{k}_{\mathbf{1}}} \mathcal{R}_{\mathbf{k}_{\mathbf{2}}} \mathcal{R}_{\mathbf{k}_{\mathbf{3}}}\right\rangle \equiv (2 \pi)^3 \delta^{(3)}\left(\mathbf{k}_{\mathbf{1}}+\mathbf{k}_{\mathbf{2}}+\mathbf{k}_{\mathbf{3}}\right) B_{\mathcal{R}}\left(k_1, k_2, k_3\right) .
\end{equation}
As can be inferred from Eq.~\eqref{eq:bispectrum}, the three momenta $\mathbf{k}_{\mathbf{1}}$, $\mathbf{k}_{\mathbf{2}}$, and $\mathbf{k}_{\mathbf{3}}$ form a triangle due to momentum conservation and we are interested in studying bispectrum's dependence on the size and shape of this triangle, i.e., its {\it geometry}.

Following the standard convention, we define the shape function $S$ as \cite{Chen_2010}\footnote{In scale-invariant models, it is common to factor out two powers of the power spectrum from the bispectrum for defining the shape function. Since, in our case, the power spectrum carries scale-dependence, we keep the full scale-dependence inside the shape function.}
,
\begin{equation}
\label{eq:shape function}
B_{\mathcal{R}}(k_1, k_2, k_3) \equiv \frac{9}{10}\,\frac{(2\pi)^4}{(k_1 k_2 k_3)^2}\,
S(k_1, k_2, k_3).
\end{equation}

Depending on the different circumstances that the inflaton field may experience during inflation, the shape function may resemble three well-known templates: local $\left(S_{\mathrm{local}} \right)$ \cite{Gangui_1994}, equilateral $\left(S_{\mathrm{eq}} \right)$ \cite{Creminelli_2006}, and folded $\left(S_{\mathrm{fold}} \right)$\cite{Meerburg_2009}. These three shapes peak near the squeezed, equilateral and folded configurations, respectively. Therefore, when compared to a given bispectrum, they serve as reliable indicators of the strength of the bispectrum at different configurations. 

All three shapes mentioned above  can be expressed in the following form,
\begin{equation}
\label{eq:S in terms of XY}
S(k_1 , k_2 , k_3) = c_x X + c_y Y + c_0,
\end{equation}
where,
\begin{equation}
\label{eq:XY definitions}
X = \frac{k_1^2}{k_2 k_3} + 2 \mathrm{perms.}, \quad Y = \frac{k_1}{k_2} + 5 \mathrm{perms}\, ,
\end{equation}
and $c_x$, $c_y$, and $c_0$ are three constants that determine the shape function. See Table \ref{tab:shapes-combination} for their numerical values for the three aforementioned shapes. Note that, the tools we will use to compare different non-Gaussianity shapes are insensitive to the overall normalization of the templates. Therefore, we have a freedom up to an overall coefficient in defining the templates. We fix this normalization such that $S_\mathrm{local} (k, k, k) = S_\mathrm{eq} (k, k, k) =1$ and $ S_\mathrm{fold} (k, k, k) = 0$.

In single-field inflation, local non-Gaussianity is produced when the inflaton deviates from its attractor solution (such as in the ultra-slow-roll (USR) inflation). The folded shape may become significant in scenarios where the initial vacuum is non–Bunch-Davies and positive and negative frequency mixing occurs. Finally, the equilateral shape is well-known to arise in models where the sound speed deviates from unity, as a result of a non-canonical kinetic term for the inflaton field  \cite{Chen_2010}.

\begin{table}[h]
\centering
\caption{The coefficients of the template defined in Eq.~\eqref{eq:S in terms of XY} for different shapes.}
\begin{tabular}{|l|ccc|}
\hline
 & $c_x$ & $c_y$ & $c_0$ \\
\hline
$S_{\text{eq}}$     & $-1$     & $1$      & $-2$   \\ 
$S_{\text{fold}}$   & $1$      & $-1$     & $3$    \\
$S_{\text{local}}$  & $\frac{1}{3}$ & $0$      & $0$    \\
$S_{\perp}$   & $-0.45$     & $0.09$      & $1.81$   \\
\hline
\end{tabular}
\label{tab:shapes-combination}
\end{table}

Generally,  $S(k_1,k_2,k_3)$ depends on both the size and the shape of the triangle formed by the three momenta $\mathbf{k}_{\mathbf{1}}$, $\mathbf{k}_{\mathbf{2}}$, and $\mathbf{k}_{\mathbf{3}}$, except for the scale-invariant scenarios which result in size-independent bispectrum (which is not of interest in this paper).
To analyze the two ways that $S$ can vary separately, it is convenient to redefine the momenta as
\begin{equation}
\label{eq:x and y}
k_1 = k, \quad k_2 = kx, \quad k_3 = ky,
\end{equation}
where $x$ and $y$ are dimensionless ratios that encode the shape of the triangle. 
Without loss of generality, we take $k$ to be the largest side of the triangle and define $x$ and $y$ as the ratios of the other sides to $k$. This choice implies $0 \leq x,y \leq 1$. It is easy to see that the template Eq.~\eqref{eq:S in terms of XY}, when rewritten in this way, is independent of $k$ due to its scale-invariance. We will see that this will not be the case for the shapes from more complex models considered in this paper.

We now turn to the tools that allow us to quantify the similarity between different shapes. Namely, we define the inner product and the shape correlator.

\subsection{Inner product and  shape correlator}
To quantitatively compare different shapes at a given comoving scale $k$, we define a $k$-dependent inner product of two shapes $S_1(k_1,k_2,k_3)$ and $S_2(k_1,k_2,k_3)$ by
\begin{equation}
\langle S_1, S_2 \rangle_k = \int_{\mathcal{V}} \mathrm{d}x \, \mathrm{d}y \; S_1(k ,k x,k y) \, S_2(k ,k x, k y) \, \mathcal{W}(k , k x, k y),
\label{eq:inner-product}
\end{equation}
where the integration domain $\mathcal{V}$ includes all $(x,y)$ in the range $0 \leq x,y \leq 1$ such that $(1,x,y)$ form a  triangle. $\mathcal{W}(k, k x, k y)$ is a weight function, which is chosen based on the specific purpose of comparing different shapes. 
We are interested in a weight function that allows us to theoretically compare the shape from an arbitrary inflationary model with the standard templates mentioned earlier, which amounts to ensuring the convergence of the integral in Eq.~\eqref{eq:inner-product}. We choose a symmetric and power-law weight function with the lowest power in momenta that ensures the IR-convergence of the dot product, especially when the local template is considered (which is, in fact, divergent in the limit $x\to 0$ or $y\to 0$):
\begin{equation}
\label{eq:weight final}
\mathcal{W}(k, kx, ky)=k_1 k_2 k_3 = k^3 x y.
% \exp\left[- a^2 k^2 \tau_*^2 (1 + x^2 + y^2)\right].
\end{equation}
Note that the inner product defined in Eq.~\eqref{eq:inner-product} can be viewed as a weighted average over different triangle configurations, ranging from the squeezed 
$\left( x \to 0, y \to 0 \right)$ to the equilateral $\left( x\to 1, y\to 1 \right)$ limits. These shapes span the perimeter range $2 k \leq P(x,y) \leq 3 k$ and the area range $0< A(x, y) \leq \frac{\sqrt{3}}{4} k^2$, where the upper (lower) bound corresponds to the equilateral (squeezed) configuration.

Using the inner product in Eq.~\eqref{eq:inner-product}, we can define the normalized shape correlator for each $k$ as,
\begin{equation}
\mathcal{C}_k(S_1, S_2) = \frac{\langle S_1, S_2 \rangle_k}{\sqrt{\langle S_1, S_1 \rangle_k \langle S_2, S_2 \rangle_k}}.
\label{eq:shape-correlator}
\end{equation}
This correlator takes values in the range $[-1,1]$ and provides a quantitative measure of how similar the two shapes are at a specific comoving scale $k$. If $|\mathcal{C}_k|$ is close to one, the correlation is complete, indicating that the two shapes share the same leading features. Conversely, if $|\mathcal{C}_k|$ is small, the two shapes behave very differently near the comoving scale $k$.

Note that while the sign of the shape correlator at a fixed scale is not physically relevant (since it changes under a sign flip of the normalization), the relative sign across different scales carries information and indicates a sign change in the amplitude of bispectrum as a function of scale.

A similar analysis based on inner products is performed in Ref.~\cite{Babich_2004}, where scale-invariant bispectra in simple inflationary models are studied. With a similar method, Ref.~\cite{Flauger:2010ja} studied inflationary models with resonance effects, where scale-invariance is broken, but the scale-dependence was not of primary interest. In this work, we focus on models that predict bispectra with both strong scale-dependence and complex shape-dependence, both of which are of our interest. That is why we define a shape correlator that depends on scale; allowing us to study the {\it geometry } of bispectrum.\footnote{
	More explicitly, the inner product defined in \cite{Babich_2004} is given by
		\begin{equation}
			\langle S_1, S_2 \rangle =\int_0^\infty \mathrm{d}k \, \int_{\mathcal{V}} \mathrm{d}x \, \mathrm{d}y \; S_1(k ,k x,k y) \, S_2(k ,k x, k y) \, \mathcal{W}(k , k x, k y),
		\end{equation}
		where, similar to the method of Ref.~\cite{Flauger:2010ja}, the integration is also performed over the scale $k$ (a feature that is also shared with).  Clearly, this approach loses the information about the scale-dependence which is also of our interest in this paper. We further note that the weight function used in Refs.\cite{Babich_2004,Flauger:2010ja} is different from ours, as a result of different motivations for defining dot products. }

It should be noted that the templates introduced earlier according to Eq.~\eqref{eq:S in terms of XY}, are not mutually orthogonal, i.e., their shape correlators are non-zero; see Table.~\ref{tab:shapes-correlators}.\footnote{Since the defined templates are scale-invariant, their shape correlator $\mathcal{C}_k$ remain $k$-independent.} As a result, caution is needed when interpreting the results. Nevertheless, a correlation close to one still indicates that the shapes being considered significantly overlap. Furthermore, if the correlation of a given shape with one template is higher than with another, it suggests a closer resemblance to the former. The non-orthogonality of the well-known templates motivates the introduction of new ones that are indeed orthogonal. We will discuss how this can be achieved next.

\begin{table}[h]
	\centering
	\caption{The correlators between different scale-invariant shapes.}
	\begin{tabular}{|l|cccc|}
		\hline
		& $S_{\text{local}}$ & $S_{\text{eq}}$ & $S_{\text{fold}}$ & $S_{\perp}$ \\
		\hline
		$S_{\text{local}}$     & $1$     & $0.72$      & $0.85$ & $0$  \\ 
		$S_{\text{eq}}$   & $0.72$      & $1$     & $0.48$  & $0.64$  \\
		$S_{\text{fold}}$  & $0.85$ & $0.48$      & $1$ & $0.0037$    \\
		$S_{\perp}$   & $0$     & $0.64$      & $0.0037$ & $1$  \\
		\hline
	\end{tabular}
	\label{tab:shapes-correlators}
\end{table}

\subsection{Orthogonal-to-Local shape}
\label{eq:orth shape}
Concerning orthogonality, a template widely discussed in the literature is the one that is orthogonal to the equilateral shape \cite{Senatore_2010}. However, since our primary interest is to study deviations from the local bispectrum, we define a shape that is orthogonal to the local template. Denoting this new template by $S_\perp$, we require
%In principle, the results obtained in the previous sections suggest a negative answer. At certain scales—particularly near the peak of the power spectrum—other configurations appear to resemble the shape of the bispectrum as much as, or even more than, the local template. 
%
%As mentioned before, the standard non-Gaussianity templates are not mutually orthogonal, i.e., their normalized shape correlators are not zero. Therefore, while a comparable or even larger shape correlator with a non-local template is meaningful, it does not necessarily imply that the bispectrum of the non-attractor models is more or equally correlated to the non-local shapes; A large part of the observed correlation could still be driven by the similarity between that template and the local one.
%
%To make a sharper statement about deviations from the local shape, we construct a new template that is explicitly orthogonal to the local one—i.e.,
\begin{equation}
	\label{eq:orth cond}
\mathcal{C}_k(S_{\rm local}, S_\perp) = 0.
\end{equation}
We use the same general template as Eq.~\eqref{eq:S in terms of XY} and find appropriate values for $c_x$, $c_y$ and $c_0$.\footnote{Since Eq.~\eqref{eq:S in terms of XY} is scale-invariant, the condition in Eq.~\eqref{eq:orth cond} holds for each $k$.} The orthogonality condition along with the normalization condition --- which we choose to be $S_\perp(k,k,k) = 1$ --- fixes two free parameters. Assuming that these two criteria determine $c_x$ and $c_0$, we arbitrarily choose $c_y =-5c_x$. This is a typical value in the sense that the resulting shape with this choice exhibits a moderate overlap with the bispectrum in the main scenario of interest, which will be discussed in Sec.~\ref{sec:transient USR}. Fine-tuned values of $\beta$ exist that show either stronger or much weaker overlap with that bispectrum which are avoided by our choice. See Table~\ref{tab:shapes-combination} for the specific choices of $c_x$, $c_y$ and $c_0$ for this shape and Table~\ref{tab:shapes-correlators} for its correlator with other templates defined earlier in this section. It is worth noting that despite the orthogonality, $S_\perp$ peaks in the squeezed configuration similar to $S_{\rm local}$.

\subsection{Projected size of non-Gaussianity}
So far, we have constructed quantities that measure the similarity between two bispectrum shapes. However, it is possible for a given bispectrum to exhibit a large correlation with a template while its amplitude at the relevant configurations is too small to be of physical significance. To address this issue, we also introduce the projected size of non-Gaussianity. We denote the size of non-Gaussianity when the shape function is projected onto a template $T$ by $f_\mathrm{NL,T}^\mathrm{proj.}$ and define it by
\begin{equation}
\label{eq:fNL}
f_\mathrm{NL,T}^\mathrm{proj.}(k) = \frac{\langle S , S_T \rangle_k }{\langle S_T , S_T \rangle_k \mathcal{P}_k^2}.
\end{equation}
For the local-type non-Gaussianity the above definition reduces to the standard $f_\mathrm{NL}$ parameter when $T$ is the local template. Therefore, the above definition is a generalization of the case of local non-Gaussianity to an arbitrary shape function projected onto an arbitrary template. However, since
different templates are not mutually orthogonal, such a quantity may be misleading in general. 
For this reason, in what follows, we restrict our use of the projected amplitude to the local and orthogonal-to-local shapes.

With these definitions and tools at hand, we are now equipped to analyze the shapes of non-Gaussianity in concrete scenarios.  
In the next section, we first consider two simple illustrative examples, before turning to the transient USR model as a physically motivated case study.

%%%%%%%%%%%%%%%%%%%%%%%%%%%%%%%%%%%%%%%%%%%%%%%%%%%%%%
\section{Preliminary Examples%: Non-Bunch--Davies USR and SR-SR transition
}
\label{sec:examples}

In order to develop a clear intuition for the analysis of the transient non-attractor models, in this section we examine two simple yet instructive examples: (i) the USR inflation with non-Bunch-Davies initial states (which we refer to by the NBD-USR model), and (ii) a model with a sharp transition between two slow-roll (SR) phases (which we refer to by the SR-SR model). These examples provide valuable insights into the behavior of more complex scenarios, such as the transient USR model which we will discuss in Sec.~\ref{sec:transient USR}.

\subsection{Example I: Ultra-Slow-Roll model with Non-Bunch--Davies Initial States}
\label{sec:NBD-USR}
Since the onset of the USR phase is expected to occur after a phase transition from an SR phase, and phase transitions tend to distort the vacuum from the Bunch-Davies (BD) state, we begin our analysis by studying a USR phase of inflation with non-BD initial conditions. 
In the USR scenario, the inflaton field rolls on a constant potential, leading to a rapid decay of the field's velocity like $\dot \phi \sim a^{-3}$. As a result, the first SR parameter $\epsilon=-\dot H/H^2$ decays as $a^{-6}$ so that the second SR parameter $\eta=\dot \epsilon/(H\epsilon)=-6$ is large but a constant. For a generic vacuum, the curvature perturbation in this model is given by
\begin{equation}
	\label{eq:mode function}
	\mathcal{R}_k(\tau)=\frac{H}{\sqrt{4 \epsilon(\tau) k^3}}\left[\alpha_k(1+i k \tau) e^{-i k \tau}+\beta_k(1-i k \tau) e^{i k \tau}\right]\, ,
\end{equation}
where $\tau$ is the conformal time and $\alpha_k$ and $\beta_k$ are the Bogoliubov coefficients that satisfy the following normalization condition:
\begin{equation}
	\label{eq:normalization}
|\alpha_k|^2 - |\beta_k|^2 = 1.
\end{equation}

We assume that the USR phase is long enough so that all modes of interest exit the horizon during the USR phase. We also assume that the transition from USR is such that the modes of interest freeze immediately after the USR phase. Since, in the USR models, non-Gaussianities are generated at superhorizon scales, the bispectrum is of the local form \cite{Namjoo_2012} (see App.~\ref{sec:cubic action} for a sketch of the derivation):
\begin{equation}
	\label{eq:NBD bispectrum}
	B_{\mathcal{R}}(k_1,k_2,k_3) =  3P_{k_1} P_{k_2}+ 2 \mathrm{perms.}\, ,
\end{equation}
where 
\begin{equation}
	P_k =\dfrac{2\pi^2}{k^3} \left(\frac{H^2}{8 \pi^2 \epsilon_e} \right)  \, |\alpha_{k} + \beta_{k}|^2\, ,
\end{equation}
where $\epsilon_e$ is the first SR parameter at the end of the USR phase. Despite its appearance, note that the shape is not necessarily local since the Bogoliubov coefficients can have complex scale-dependence. 

%As discussed in App.~\ref{sec:cubic action}, in USR models the dominant contribution to non-Gaussianity originates from field redefinition terms \cite{Namjoo_2012}. The bispectrum for USR models with a non-Bunch--Davies (NBD) initial state can be expressed as
%\begin{equation}
%\label{eq:NBD bispectrum}
%B_{\mathcal{R}}(k_1,k_2,k_3) = 12 \pi^4 \left(\frac{H^2}{8 \pi^2 \epsilon_e} \right)^2 \left( \frac{|\alpha_{k_1} + \beta_{k_1}|^2 |\alpha_{k_2} + \beta_{k_2}|^2 }{k_1^3 k_2^3} + 2 \mathrm{perms.} \right),
%\end{equation}
%where $\alpha_k$ and $\beta_k$ are the Bogoliubov coefficients encoding the deviation from the Bunch--Davies vacuum, $H$ is the Hubble parameter, and $\epsilon_e$ is the first SR parameter at the end of the USR phase, defined as $\epsilon = -\dot{H}/H^2$.

%If $\alpha_k$ and $\beta_k$ are constant, coherent, and independent of the wavenumber, the resulting shape would be local. However, in more realistic cases a transition modifies the vacuum state, and hence the Bogoliubov coefficients become scale-dependent and the shape would be non-trivial.

\begin{figure}[t]
    \centering
    \includegraphics[width=0.45\textwidth]{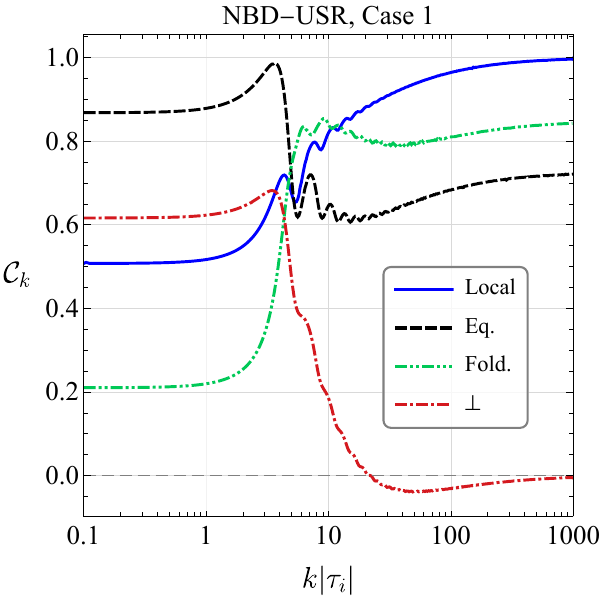}
        \hspace{.5cm}
    \includegraphics[width=0.45\textwidth]{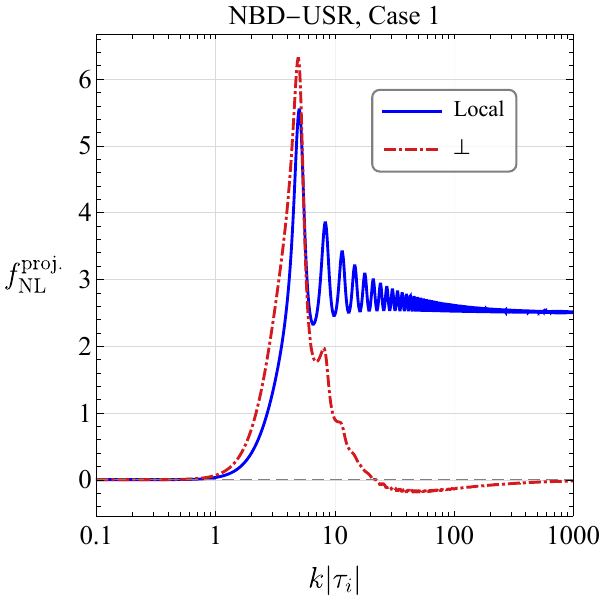}
    \caption{
       The shape correlator (left) and the projected amplitude of the local and orthogonal-to-local shapes (right) for the non-Bunch-Davies USR model (Case 1), descibed in Eq.~\eqref{eq:NBD USR case1}. The power spectrum peaks around $k|\tau_i| \simeq 3$.
%        The shape correlator for the non-Bunch-Davies USR model (Case 1), descibed in Eq.~\eqref{eq:NBD USR case1}. 
%        \textit{Left panel:} correlation with the standard local, equilateral, and folded templates as a function of scale.  
%        \textit{Right panel:} projected amplitudes $f_{\mathrm{NL}}^{\mathrm{local}}$ and $f_{\mathrm{NL}}^{\perp}$ as defined in Sec.~\ref{sec:Tools}.  
%        On large scales, the bispectrum resembles the equilateral template due to the sharp transition, while in the short scales it approaches the local template with amplitude $f_{\mathrm{NL}}^{\mathrm{local}} \to 2.5$, as the model approches the USR model with Bunch-Davies vacuum.  
%        Around the transition scale, the projected amplitudes of the local and orthogonal shapes become comparable, highlighting the importance of considering the full-shape bispectrum.
%{\bf Isn't it better to use $k|\tau_i|$ rather than $-k\tau_i$?} 
   }
    \label{fig:case1}
\end{figure}

The explicit form of the Bogoliubov coefficients depends on the details of the phase transition. In this subsection, we explore a few illustrative choices.  
First, we choose the Bogoliubov coefficients to be those obtained from an infinitely sharp transition between an initial SR phase and a subsequent USR phase (see App.~\ref{sec:mode functions} and Ref.~\cite{Cai_2018} for the full derivation),
\begin{equation}
\label{eq:NBD USR case1}
 \text{Case 1}:\quad   \beta_k = -\frac{3 i (1+i k \tau_i)^2}{2 k^3 \tau_i^3} e^{-2 i k \tau_i}, \quad \alpha_k = 1 + \frac{3 i \left(k^2 \tau_i^2+1\right)}{2 k^3 \tau_i^3}\, ,
\end{equation}
where $\tau_i$ denotes the transition conformal time.\footnote{In a realistic model, the modes experience a phase of SR inflation before the transition, in which case, our results will be invalid for scales satisfying $k|\tau_i| <1$. However, here we use Eq.~\eqref{eq:NBD USR case1} as given for all modes and assume USR is the only relevant phase of inflation. The more complex --- and more realistic --- model will be studied in Sec.~\ref{sec:transient USR}.}
%
%%The shape correlator of this case with the standard non-Gaussianity templates, together with the projected amplitudes of the local and orthogonal-to-local shapes, is shown in Fig.~\ref{fig:case1}.  
%%As can be seen, the shape of the USR model with a non-Bunch–Davies vacuum—induced by a sharp transition—resembles the equilateral template on large scales, reflecting the impact of the transition.  
%%Moreover, the Bogoliubov coefficients become trivial in the short scale limit, i.e. $\alpha_k \to 1$ and $\beta_k \to 0$, and the bispectrum smoothly approaches that of the USR model with a Bunch–Davies vacuum.  
%%Thus, in this regime, the shape increasingly resembles the local template, with an amplitude approching $f_{\mathrm{NL, local}}^{\mathrm{proj.}} = 2.5$, as it is well-known.  
%
The corresponding shape correlator and projected amplitudes are shown in Fig.~\ref{fig:case1}. On large scales, the bispectrum resembles the equilateral template, reflecting the impact of the transition. In the short-scale limit, the Bogoliubov coefficients approach BD values ($\alpha_k \to 1$, $\beta_k \to 0$), and the bispectrum smoothly reduces to that of USR inflation with a BD vacuum, reproducing the well-known local result $f_{\mathrm{NL, local}}^{\mathrm{proj.}} \to 2.5$ \cite{Namjoo_2012}. Interestingly, the projected amplitudes $f_{\rm NL, local}^{\rm proj.}$ and $f_{\rm NL, \perp}^{\rm proj.}$ exhibit comparable peaks around the same  scale. Importantly, the peak of the power spectrum is located around $k|\tau_i| \simeq 3$ where the bispectrum has significant overlap with the equilateral template.

%This behavior is clearly visible in the right panel of Fig.~\ref{fig:case1}, where the projected local amplitude grows from a small value and converges to the limiting value $f_{\rm NL, local}^{\rm proj.} = 2.5$.  
%Interestingly, the projected amplitudes $f_{\rm NL, local}^{\rm proj.}$ and $f_{\rm NL, \perp}^{\rm proj.}$ exhibit comparable peaks around the transition scale.

\begin{figure}[t]
    \centering
    \includegraphics[width=0.45\textwidth]{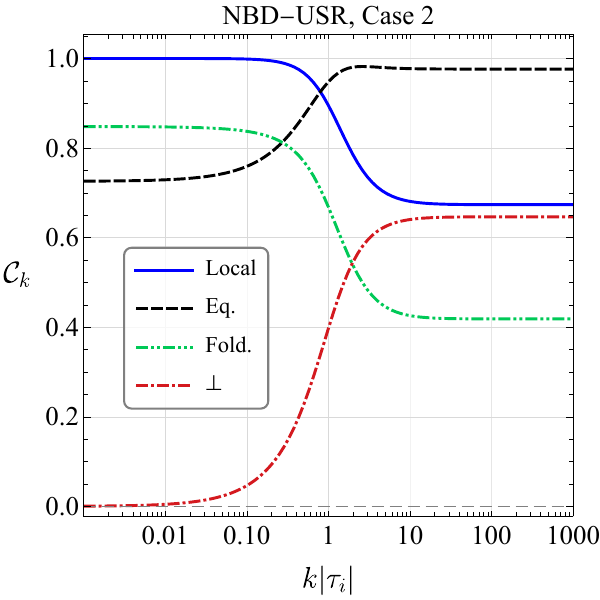}
        \hspace{.5cm}
    \includegraphics[width=0.45\textwidth]{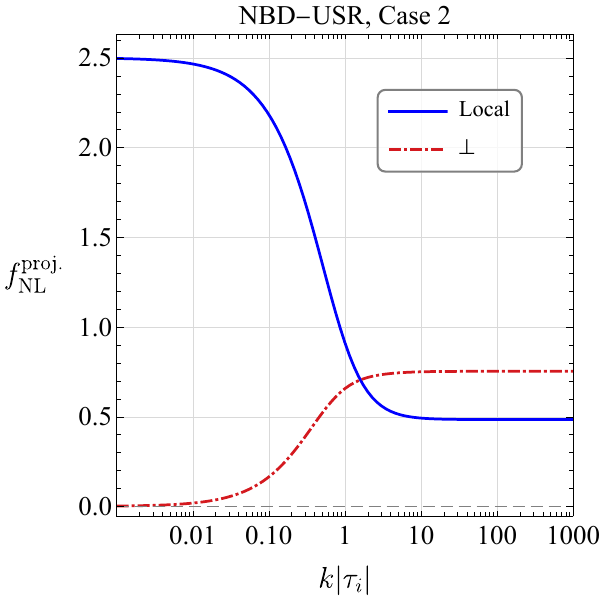}
	\caption{
	The shape correlator (left) and the projected amplitude of the local and orthogonal-to-local shapes (right) for the NBD-USR model (Case 2), descibed in Eq.~\eqref{eq:NBD USR case2}. 
%The shape correlator for the non-Bunch-Davies USR model (Case 2), descibed in Eq.~\eqref{eq:NBD USR case2}. 
%        \textit{Left panel:} correlation with the standard local, equilateral, and folded templates as a function of scale.  
%        \textit{Right panel:} projected amplitudes $f_{\mathrm{NL, local}}^{\rm proj.}$ and $f_{\mathrm{NL}, \perp}^{\rm proj.}$ as defined in Sec.~\ref{sec:Tools}.  
%        Whenever particle production occure, the bispectrum becomes more correlated with the equilateral template and $f_{\rm NL}^\perp$ grows accordingly. 		        Thus the non-local contributions to the bispectrum increases with the particle production. 
%       {\bf Please add - between NBD and USR. The font can still be a bit smaller. In the legends, use Eq. and Fold. (add a dot). Can the aspect ratio of the two figures be the same?}
        }
    \label{fig:case2}
\end{figure}

\begin{figure}[t]
    \centering
    \includegraphics[width=0.45\textwidth]{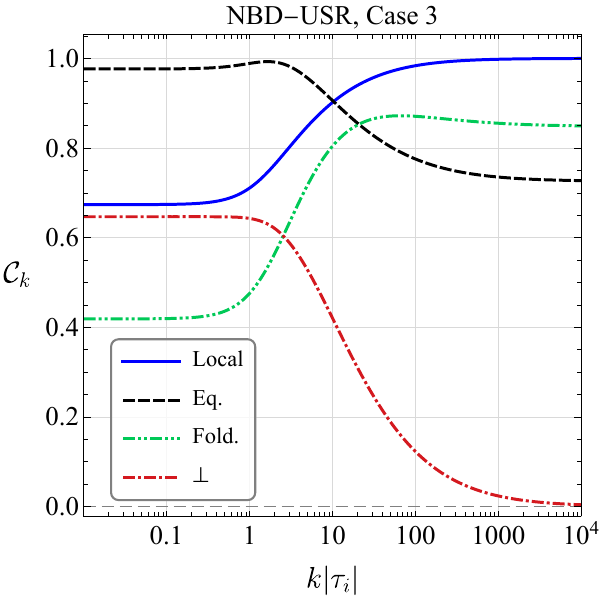}
    \hspace{.5cm}
    \includegraphics[width=0.45\textwidth]{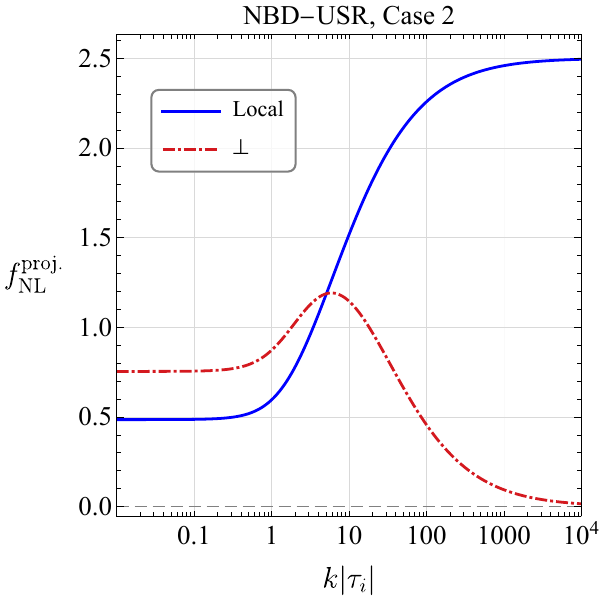}
    \caption{
    The shape correlator (left) and the projected amplitude of the local and orthogonal-to-local shapes (right) for the NBD-USR model (Case 3), descibed in Eq.~\eqref{eq:NBD USR case3}. 
%   	 The shape correlator for the non-Bunch-Davies USR model (Case 3), descibed in Eq.~\eqref{eq:NBD USR case3}. 
%        \textit{Left panel:} correlation with the standard local, equilateral, and folded templates as a function of scale.  
%        \textit{Right panel:} projected amplitudes $f_{\mathrm{NL, local}}^{\rm proj.}$ and $f_{\mathrm{NL}, \perp}^{ \rm proj.}$ as defined in Sec.~\ref{sec:Tools}.  
%        As in the previous case, particle production drives the bispectrum towards the equilateral shape, with the orthogonal amplitude growing accordingly. The 				figure highlights that, in this scenario, the non-local contributions can significantly exceed the local one. 
}
    \label{fig:case3}
\end{figure}

Next, we examine two scenarios where particle production, likely caused by phase transitions, occurs either at small scales or at large scales. Since the occupation number is proportional to $|\beta_k|^2$ we model these two cases by adopting simple forms for $\beta_k$ and determine $\alpha_k$ from the  normalization condition Eq.~\eqref{eq:normalization}. An  arbitrary relative phase remains undetermined which we then fix by requiring a finite power spectrum at large scales. We then have
\begin{eqnarray}
\label{eq:NBD USR case2}
& \text{Case 2}:& \quad \beta_k = - k \tau_i,\quad \alpha_k = \sqrt{1 + k^2 \tau_i^2},
\\
\label{eq:NBD USR case3}
 &\text{Case 3}:& \quad \beta_k = \frac{-1}{k \tau_i}, \quad \alpha_k = \frac{1}{k \tau_i} \sqrt{1 + k^2 \tau_i^2}.
\end{eqnarray}
Case 2 and Case 3 correspond to particle production at short and large scales, respectively. We do not attempt to justify the above choices through a detailed analysis of a specific model; instead, we consider them as toy examples that illustrate the effect of particle production on the bispectrum. 
Figs.~\ref{fig:case2} and \ref{fig:case3} show the shape correlators and projected amplitudes for these two cases. In both cases, the bispectrum at scales most affected by particle production has a strong overlap with the equilateral template.\footnote{This observation suggests a new way to generate (scale-invariant) equilateral-type non-Gaussianity, distinct from its production due to non-canonical kinetic term, which deserves further study: continuous particle production during inflation.} Conversely, in the opposite regime, the overlap with the local shape increases due to the USR inflation, which tends to produce local non-Gaussianity. In this limit, we recover the well-known size of non-Gaussianity $f_{\mathrm{NL, local}}^{\mathrm{proj.}} \to 2.5$.

\subsection{Example II: Sudden Transition Between Two Slow-Roll Phases}
\label{sec:SR-SR}
%{\bf I suggest we make the following replacements to avoid confusion with the next section: $\epsilon_i \to \epsilon_1, \epsilon_V\to \epsilon_2, h\to \kappa, V_0 \to V_c, \phi_i \to \phi_c, \tau_i \to \tau_c$.} \bnote{Do you still insist on this?}

As another illustrative example, we now investigate a scenario in which the inflaton field experiences a sharp transition due to a break in the potential. This results in a short period of non-attractor evolution after the transition until the field relaxes to the new attractor. To be more concrete, we consider an effective potential of the form
%,\footnote{This model can be viewed as a special limit of the transient USR model in which the two ends of the USR interval are brought infinitesimally close together. See App.~\ref{sec:background} for more details.}
\begin{equation}
\label{eq:SR-SR Potential}
V(\phi)=V_0\left[1+\sqrt{2 \epsilon_V}\left(\phi-\phi_i\right) \theta\left(\phi_i-\phi\right)+\sqrt{2 \epsilon_i}\left(\phi-\phi_i\right) \theta\left(\phi-\phi_i\right)\right],
\end{equation}
where $V_0$ is a constant and $\epsilon_i$ and $\epsilon_V$ quantify the slope of the potential before and after the break, respectively. We denote the transition conformal time by $\tau_i$ when the field reaches $\phi_i$. Note that the first SR parameter $\epsilon$ is approximately constant before and well after the transition and is approximately equal to $\epsilon_i$ and  $\epsilon_V$, respectively. However, it can  evolve significantly during the relaxation period.

The details of the background solution of this model are discussed in App.~\ref{sec:background}. Here we only introduce the  main parameter that quantifies the physics of the transition, which we denote by $\hat h$ and is defined as
\begin{equation}
\label{eq:h SR SR}
\hat h = -6 \sqrt{\frac{\epsilon_V}{\epsilon_i}}.
\end{equation}
%where $\epsilon_V$ is the first SR parameter of the second SR phase.
Note that $\hat h < 0$, and the larger the $|\hat h|$, the faster the relaxation occurs. If $\hat h=-6$, there is no transition and the relaxation period disappears. In the limit $\hat h\to -\infty$, the relaxation period is long but the curvature perturbation freezes immediately after the transition. In all numerical computations presented in this section, we set $\hat h = -1000$.

\begin{figure}[t]
    \centering
    \includegraphics[width=0.45\textwidth]{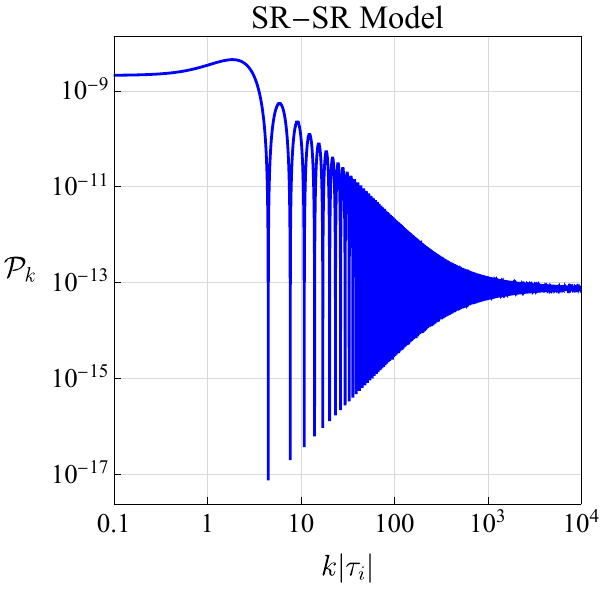}
    \caption{
    The power spectrum for the SR-SR model. In this plot, we have set $\hat h = -1000$, $\epsilon_i = 10^{-3}$ and chose $H$ such that the power spectrum at the largest scale matches its observed value on CMB scales, i.e. $\mathcal{P}_k^\mathrm{CMB} \approx 2.1 \times 10^{-9}$.
    }
    \label{fig:SR-SR power}
\end{figure}

The mode function in this scenario is still given by Eq.~\eqref{eq:mode function}. Before the transition, during the first SR phase, we have $\alpha_k = 1$ and $\beta_k = 0$ due to the BD vacuum. In the second phase, the Bogoliubov coefficients  are determined by imposing the continuity of $\mathcal{R}$ and $\mathcal{R}'$ and are given by Eq.~\eqref{eq:NBD USR case1}. 
%\begin{equation}
%\label{eq:mode function}
%   \mathcal{R}_k(\tau)=\frac{H}{\sqrt{4 \epsilon(\tau) k^3}}\left[\alpha_k(1+i k \tau) e^{-i k \tau}+\beta_k(1-i k \tau) e^{i k \tau}\right]
%\end{equation}
%where $\alpha_k$ and $\beta_k$ are the Bogoliubov coefficients. In the first SR phase, these coefficients are trivial due to the Bunch--Davies vacuum, i.e. $\alpha_k = 1$ and $\beta_k = 0$, while in the second phase they are determined by imposing continuity of $\mathcal{R}$ and $\mathcal{R}'$. 
See App.~\ref{sec:mode functions} for further details and Fig.~\ref{fig:SR-SR power} for the behavior of the power spectrum for this model. 

Once the mode function is known, the bispectrum can be obtained by direct calculations using the in-in formalism. As it is argued in App.~\ref{sec:cubic action}, the dominant contribution to the three-point function in this model arises from the term $S_3 \supset \int d \tau d^3 x \frac{a^2 \epsilon}{2} \eta^{\prime} \mathcal{R}^2 \mathcal{R}^{\prime}$ in the cubic action, where $\eta \equiv \dot{\epsilon} / H \epsilon$ is the second SR parameter and prime denotes the derivative with respect to conformal time. The bispectrum receives two distinct contributions, one associated with the transition and another with the relaxation to the second SR attractor. We leave the technical details to App.~\ref{sec:bispectrum} and present here the final expression for the bispectrum:
\begin{equation}
\label{eq:SR SR bispectrum}
B_{\mathcal{R}}(k_1,k_2,k_3) = \frac{2(h+6) \epsilon _i}{H^2 \tau _i^3}\text{Im}\left\{P_{k_1}\left(\tau _i, 0\right)P_{k_2}\left(\tau _i, 0\right) P_{k_3} \left(\tau _i,0\right) \left(\frac{h}{2} + \Gamma_{k_1} +\Gamma_{k_2} +\Gamma_{k_3}  \right)\right\}\, ,
\end{equation}
where 
\begin{equation}
	\label{eq:Gamma}
	\Gamma_k \equiv \frac{ k^2 \tau _i^2}{1-i k \tau _i}.
\end{equation}

%The bispectrum for this model will be as follows,
%\begin{equation}
%B_{\mathcal{R}}(k_1,k_2,k_3) = \frac{(h+6) \epsilon _i}{H^2 \tau _i^3}\text{Im}\left\{P_{k_1}\left(\tau _i, 0\right)P_{k_2}\left(\tau _i, 0\right) P_{k_3} \left(\tau _i,0\right) \left(h +\frac{2 k_1^2 \tau _i^2}{1-i k_1 \tau _i} + \text{perms}. \right)\right\},
%\end{equation}
%where $\tau_i$ is the conformal time at $\phi_i$, and the relaxation parameter, $h$, is defined as,
%\begin{equation}
%\label{eq:h SR SR}
%h = -6 \sqrt{\frac{\epsilon_V}{\epsilon_i}}
%\end{equation}

\begin{figure}[t]
    \centering
    \includegraphics[width=0.45\textwidth]{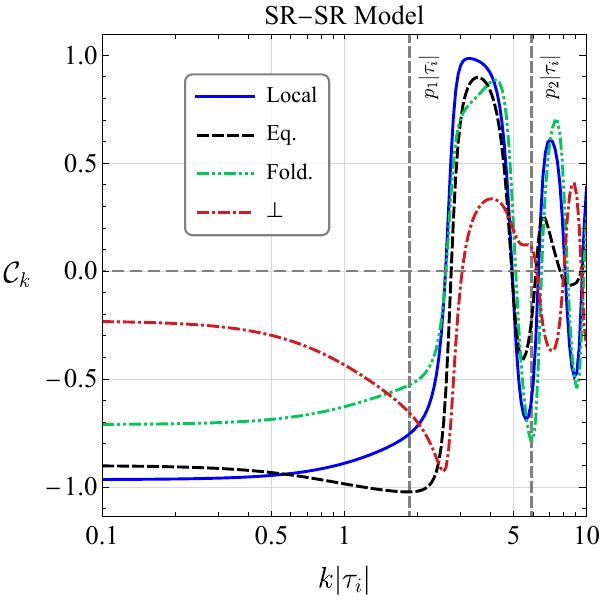}
        \hspace{.5cm}
    \includegraphics[width=0.45\textwidth]{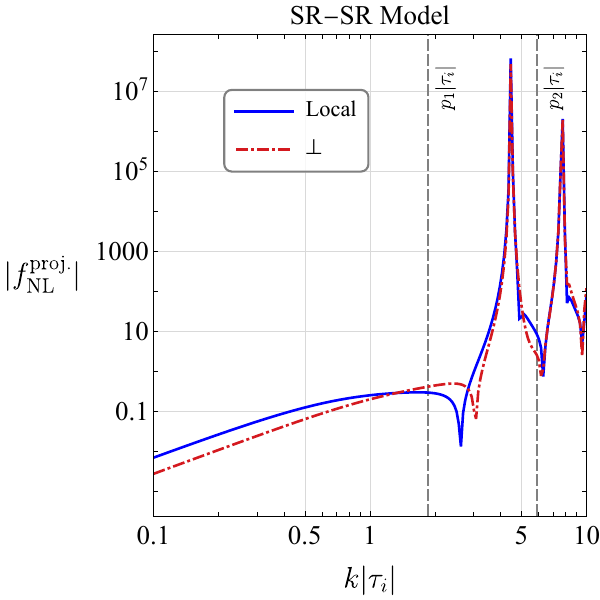}
    \caption{
    The shape correlator (left) and the projected amplitude of the local and orthogonal-to-local shapes (right) for the SR-SR  model. The choice of parameters is the same as in Fig.~\ref{fig:SR-SR power}. The vertical dashed lines show the scale of the power spectrum’s highest and second-highest peaks, denoted by $p_1$ and $p_2$, respectively. 
    }
    \label{fig:SR-SR}
\end{figure}

 Fig.~\ref{fig:SR-SR} shows the shape correlators and the projected amplitudes for this model. On large scales, the amplitude of the bispectrum vanishes, as is expected from the results of SR inflation and the SR consistency relation \cite{Maldacena_2003, Namjoo:2023rhq, Namjoo:2024ufv}.\footnote{Since many terms in the cubic action that lead to small non-Gaussianities are neglected, we can only assess the smallness of the bispectrum at the largest scales rather than obtaining the full result.}  (The transition is not visible to these scales since the curvature perturbation is frozen during the transition.) The first non-zero contribution at largest scales is given by 
\begin{equation}
\label{eq:small scale SR SR}
B_{\mathcal{R}}(k_1, k_2, k_3) \simeq  \frac{6+h}{5h} \left(\frac{H^2}{4 \epsilon_i k_1 k_2 k_3}\right)^2 (k_1 + k_2 + k_3)^2\, \tau_i^2 \left[X - 2Y + 6 \right], \quad \text{for $k_j\tau_i \to 0$}\, ,
\end{equation}
where $j=\{1,2,3 \}$ and $X$ and $Y$ are defined in Eq.~\eqref{eq:XY definitions}.
This bispectrum is highly correlated with the local template and exhibits the same form of IR-divergence in the squeezed limit. In fact, in the limit $k_j\tau_i \to 0$ for all momenta and $k_1\ll k_2 \sim k_3$, the bispectrum behaves like $B_{\mathcal{R}} \sim k_3^2 \tau_i^2\, \times  (k_3 k_1)^{-3}$. The first factor shows the amplitude of the squeezed limit bispectrum vanishes at large scales ($k_3\tau_i \to 0$) while the second factor is the usual IR divergence of local non-Gaussianity, both of which are consistent with Fig.~\ref{fig:SR-SR}. % The effective non-Gaussianity amplitude also goes to zero as the scale increases. This is consistent with intuition: modes much larger than the transition scale are insensitive to the localized feature in the potential.

At shorter scales, the sharp transition leads to particle production and  induces strong correlations among the modes that leave the horizon around the transition time, causing the bispectrum to have large (but negative) correlation with the equilateral shape.  Interestingly, the peak of the power spectrum, the scale of interest for the PBH and SIGW production, lies in this regime. In Fig.~\ref{fig:SR-SR} this scale is identified by a vertical line and is denoted by $p_1$. For comparison, we also show the scale of the second-highest peak, denoted by $p_2$.

Finally, at much smaller scales, both the power spectrum and the bispectrum exhibit significant oscillations, due to deviations from the BD vacuum. Consequently, the correlations with all scale-invariant templates of non-Gaussianity also oscillate, and none of the considered templates dominate the shape correlator in this regime.

%For this model, the full shape correlators take the following form,
%\begin{equation}
%    \mathcal{C}^\mathrm{Eq.} = ,\quad \mathcal{C}^\mathrm{Local} = ,\quad \mathcal{C}^\mathrm{Fold} = 
%\end{equation}
%
%[Explain!]
%
%Moreover, the projected non-Gaussianity amplitudes for the local and orthogonal-to-local shapes are given by.
%\begin{equation}
%f_{\rm NL}^{\rm local} = ,\quad f_{\rm NL}^\perp = 
%\end{equation}
%
%[Explain!]

Let us summarize the lessons we learned from the two examples explored in this section. In a pure non-attractor model, the bispectrum is expected to have a local form as given in Eq.~\eqref{eq:NBD bispectrum}. However, due to phase transitions to and from the non-attractor phase, the actual shape can deviate significantly from local, except for the modes that exit the horizon much earlier than the transition time, for which the evolution resulting from the transition remains scale-invariant.   Around the transition scales --- where the particle production is most effective and where a peak in the power spectrum is expected to appear --- the bispectrum is strongest near the equilateral configuration. 

Since phase transitions are expected to cause deviations from the BD vacuum, a question remains: why does the bispectrum not show a strong correlation with the folded shape, which is expected to be generated from positive and negative frequency mixing, both of which are present in the non-BD vacuum? For the scenarios under consideration, this mixing is not expected to efficiently occur for the following reasons. We have seen that two sources of large non-Gaussianity exist in the explored models. First, correlations generated by phase transitions, which occur quickly and do not allow significant mixing. Second, non-attractor phases, where strong correlations mainly occur at large and superhorizon scales, where oscillations with positive and negative frequencies disappear due to the classicalization process. Therefore, in these scenarios, neither of these sources can generate folded non-Gaussianity, despite the large deviations from the BD vacuum.

With the intuition obtained in this section, we now study the transient USR model which is a popular model for producing PBHs and SIGWs.
\section{The Transient Ultra-Slow-Roll Model}
\label{sec:transient USR}
%In this section, we briefly review the transient ultra-slow-roll (USR) inflation, and then derive its bispectrum and examine its correlation with the non-Gaussianity templates reviewed in Sec.~\ref{sec:Shape}.

The transient USR model consists of a USR phase that is sandwiched between two SR phases. The potential for this model can be effectively described as,
\begin{equation}
\label{eq:Potential}
V(\phi)=V_0\left[1+\sqrt{2 \epsilon_V}\left(\phi-\phi_e\right) \theta\left(\phi_e-\phi\right)+\sqrt{2 \epsilon_i}\left(\phi-\phi_i\right) \theta\left(\phi-\phi_i\right)\right],
\end{equation}
where the USR phase lasts from $\phi_i$ to $\phi_e$, and $\epsilon_i$ and $\epsilon_V$ quantify the slopes of the potential in the first and last SR phases respectively. This model reduces to the SR-SR model, studied in Sec.~\ref{sec:SR-SR}, in the limit $\phi_e \to \phi_i$ where the USR phase disappears.

This model is well-known and the details of the background solution and the linear perturbations can be found e.g., in Refs.~\cite{Namjoo:2024ufv, firouzjahi2023oneloopcorrectionspowerspectrum}. Some details are also discussed in App.~\ref{sec:details}. Here, we only briefly discuss the physics behind this model. 

The inflaton field rolls in the first attractor phase, characterized by the first SR parameter $\epsilon_i $, until it reaches $ \phi_i $, at  conformal time $ \tau_i$. It then transits instantly to the flat segment of the potential that lies between $ \phi_i $ and $ \phi_e $, where the USR phase occurs. 
During the USR phase, $\eta = -6$ and the first SR parameter decays like $ \epsilon \sim a^{-6}$. Thus, once the inflaton field reaches $ \phi_e $, at conformal time $ \tau_e$, the first SR parameter reaches the value $\epsilon_e = e^{-6\Delta N} \epsilon_i$, where $\Delta N=\ln(\tau_i/\tau_e)$ is the number of efolds during the USR phase \cite{Namjoo_2012}.
Subsequently, the inflaton field instantaneously starts rolling on the second SR segment of the potential and eventually relaxes to the final attractor with the first SR parameter $\epsilon_V$. The relaxation parameter for this model is defined by 
 \begin{eqnarray}
 h = -6 \sqrt{\frac{\epsilon_V}{\epsilon_e}}\, ,
 \end{eqnarray}
 which reduces to $\hat h$ defined in Eq.~\eqref{eq:h SR SR} for the SR-SR  model in the limit $\tau_e\to \tau_i$ (or $\epsilon_e \to \epsilon_i$). As in the SR-SR  model, depending on the value of $h$, the field either accelerates or decelerates until it relaxes to the second attractor.

To track the curvature perturbations, we again use the general form of the mode function presented in Eq.~\eqref{eq:mode function}. The Bogoliubov coefficients before the first transition are given by the BD vacuum, while in the second and third phases they are determined by imposing the continuity of $\mathcal{R}_k$ and $\mathcal{R}_k^\prime$ at the transitions. Further details are provided in App.~\ref{sec:mode functions}. 

\begin{figure}[h!]
\centering
\includegraphics[width=0.45\textwidth]{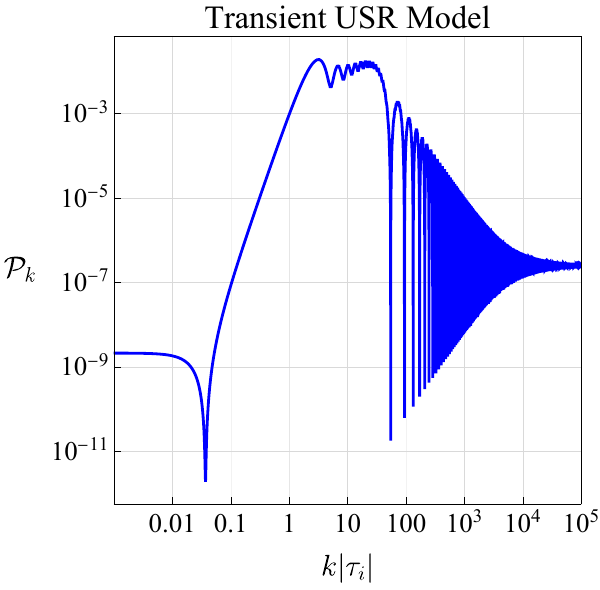}
\caption{
Power spectrum of the transient USR model. For this plot, we have set $h = -1000$, $\Delta N \equiv \ln(\tau_i/\tau_e)=2.5$, $\epsilon_i = 10^{-3}$ and chose $H$ such that the power spectrum at the largest scales matches its observed value, i.e. $\mathcal{P}_k^\mathrm{CMB} = 2.1 \times 10^{-9}$.}
\label{fig:power-spectrum}
\end{figure}

The behavior of the final power spectrum for a specific choice of parameters is demonstrated in Fig.~\ref{fig:power-spectrum}. We observe that, unlike the SR-SR  model, there are two comparable peaks caused by the presence of two transitions in this model. For our choice of parameters, the power spectrum at the highest and second-highest peaks are related by ${\cal P}_{p_2}\simeq 0.9 {\cal P}_{p_1}$, where $p_1$ and $p_2$ denote the comoving scales of the highest and second-highest peaks, respectively. It is commonly assumed that the highest peak is responsible for the PBH production, which naively seems reasonable since the PBH abundance is very sensitive to the power spectrum. However, as will be discussed in Sec.~\ref{sec:beta}, when the complications associated with non-Gaussianities are taken into account, this assumption will be challenged. Note that this result matters not only  for the PBH abundance but also for their mass since the two peaks occur at very different scales; in this example, they are related by $p_2 \simeq 6.3 p_1$. The same complication is expected to arise for SIGWs.

The three-point function for this model can be obtained through a direct in-in calculation. One can show that the dominant contribution still arises from $S_3 \supset \int d \tau d^3 x \frac{a^2 \epsilon}{2} \eta^{\prime} \mathcal{R}^2 \mathcal{R}^{\prime}$ in the cubic action (see App.~\ref{sec:cubic action}). Since this terms is proportional to $\eta'$, the bispectrum receives three significant contributions, from the first transition, the second transition, and the phase of relaxation (to the second SR inflation); which we denote them by  $B_\mathcal{R}^{T_1}, B_\mathcal{R}^{T_2}$, and $B_\mathcal{R}^{R}$, respectively.  The USR phase does not contribute since $\eta$ is a constant in this period. We have
(see App.~\ref{sec:bispectrum} for further details)
\begin{equation}
\label{eq:transient USR bT1}
B_\mathcal{R}^{T_1} = \frac{12 \epsilon_i}{H^2 \tau_i^3} \mathrm{Im}\big[ P_{k_1}(\tau_i , 0) P_{k_2}(\tau_i , 0) P_{k_3}(\tau_i , 0) \big(  \Gamma_{k_1} +\Gamma_{k_2} +\Gamma_{k_3} \big) \big], 
\end{equation}
\begin{equation}
\label{eq:transient USR bT2}
B_\mathcal{R}^{T_2} = \frac{2 h \epsilon_e}{H^2 \tau_e^3} \mathrm{Im}\left[ P_{k_1}(\tau_e , 0) P_{k_2}(\tau_e , 0) P_{k_3}(\tau_e , 0) \left( -9 + \Lambda_{k_1} + \Lambda_{k_2}+\Lambda_{k_3}\right) \right],
\end{equation}
\begin{equation}
\label{eq:transient USR bR}
B_\mathcal{R}^{R} = \frac{h(h+6) \epsilon_e}{H^2 \tau_e^3} \mathrm{Im} \left[ P_{k_1}(\tau_e , 0) P_{k_2}(\tau_e , 0) P_{k_3}(\tau_e , 0) \right],
\end{equation}
where $\Gamma_k$ is defined in Eq.~\eqref{eq:Gamma} and
%$\epsilon_e$ is the first slow-roll parameter evaluated at the end of the USR phase, $\epsilon_e = \left( \frac{\tau_e}{\tau_i}\right)^6 \epsilon_i$, and $ h = -6 \sqrt{\frac{\epsilon_V}{\epsilon_e}}$ is the relaxation parameter and 
$\Lambda_k$ is defined by
\begin{equation}
\label{eq:Lambda}
\Lambda_k \equiv \frac{k^2 \tau_e^2 \left(\alpha_k^* e^{i k \tau_e} + \beta_k^* e^{-i k \tau_e} \right)}{\alpha_k^* (1- i k \tau_e) e^{i k \tau_e} + \beta_k^* (1+ i k \tau_e) e^{- i k \tau_e}},
\end{equation} 
where $\alpha_k$ and $\beta_k$ are the Bogoliubov coeficients in the USR phase and are still given by Eq.~\eqref{eq:NBD USR case1}.
The total bispectrum is then given by
\begin{equation}
	\label{eq:transient USR bispectrum}
	B_\mathcal{R} = B_\mathcal{R}^{T_1} + B_\mathcal{R}^{T_2} + B_\mathcal{R}^R\, .
\end{equation}

Note that Eq.~\eqref{eq:transient USR bispectrum} represents the full shape bispectrum for the transient USR model, valid for all configurations and scales. To the best of our knowledge, this is the first time the full bispectrum has been computed for this model, including the effect of the relaxation period.\footnote{We note that Ref.~\cite{Tasinato:2023ioq} computes the bispectrum in the large $\eta$ limit of the non-attractor phase, assuming the bispectrum arises solely from the phase transitions.} However, it is straightforward to verify that in the squeezed limit, Eq.~\eqref{eq:transient USR bispectrum} reduces to the results obtained in Ref.~\cite{Namjoo:2024ufv} using consistency relations.

\begin{figure}[h!]
\centering
  \includegraphics[width=0.45\textwidth]{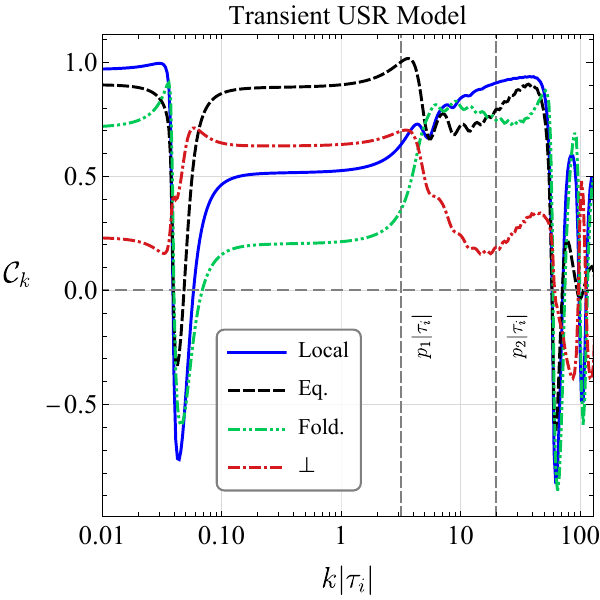}
      \hspace{.5cm}
\includegraphics[width=0.45\textwidth]{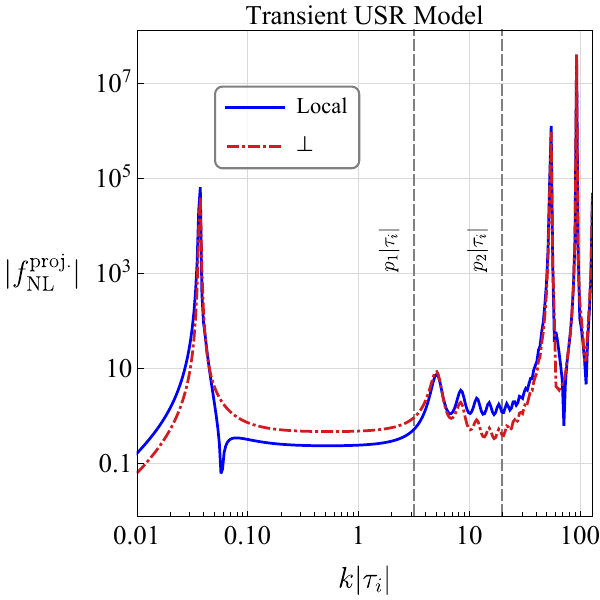}
\caption{
 The shape correlator (left) and the projected amplitudes $f_{\mathrm{NL, local}}^{\rm proj.}$ and $f_{\mathrm{NL}, \perp}^{ \rm proj.}$ (right) for the transient USR model. The scale of the highest and second-highest peaks of the power spectrum are denoted by $p_1$ and $p_2$, respectively and are highlighted by the vertical dashed lines. The choice of parameters is the same as in Fig.~\ref{fig:power-spectrum}.}
\label{fig:TUSR}
\end{figure}

Having found the analytic expression for the full bispectrum, we are prepared to study its {\it geometry}. Fig.~\ref{fig:TUSR}  shows the correlation of  this bispectrum with the well-known non-Gaussianity templates along with the projected amplitudes of non-Gaussianity. Again, we observe that the bispectrum significantly overlaps with different templates at different scales.\footnote{Note that this is very different from resonant non-Gaussianity \cite{Flauger:2010ja}. Despite both models predicting oscillatory behavior of the bispectrum, resonant non-Gaussianity shows almost no overlap with any of the templates we considered here.}
Similar to the SR-SR model, in the large scale limit, the shape of the bispectum is local with vanishing amplitude. In fact, in the limit where all momenta are much smaller than $1/|\tau_i|$, the bispectrum is given by Eq.~\eqref{eq:small scale SR SR}, up to a numerical factor.

As we move to higher momenta, the non-Gaussianity generated by the first transition becomes negligible. However, the first transition still influences the mode function leading to deviations from the BD vacuum. Therefore, in this regime, the bispectrum effectively resembles that of a USR model with a non-BD vacuum, discussed in Sec.~\ref{sec:NBD-USR}. This similarity can be clearly observed by comparing the left panel of Fig.~\ref{fig:TUSR} in the range $0.1<k|\tau_i|<20$ with the left panel of Fig.~\ref{fig:case1}. The peak scale of the power spectrum $p_1$ lies in the regime where the bispectrum shows an almost complete correlation with the equilateral shape. This indicates that the equilateral configuration is in fact more important at the highest peak than the squeezed configuration. 

At very short scales, the bispectrum begins to oscillate and loses significant correlation with the standard scale-invariant templates. In this regime, the power spectrum itself has decayed to negligible values, making observational consequences insignificant.

\begin{figure}[h!]
	\centering
	\includegraphics[width=0.45\textwidth]{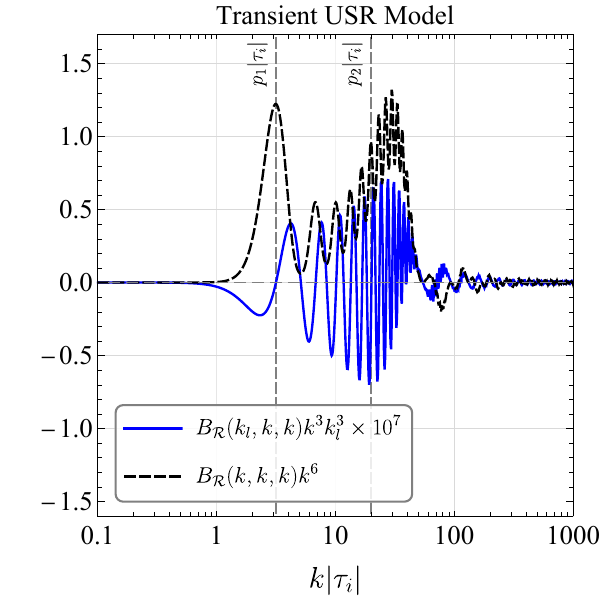}
	\caption{
		The bispectrum of the transient USR model in the squeezed ($k_l \ll k$) and equilateral limits. The choice of parameters is the same as in Fig.~\ref{fig:power-spectrum}.  The scale of the highest and second-highest peaks of the power spectrum are denoted by $p_1$ and $p_2$, respectively and are highlighted by the vertical dashed lines. 
	}
	\label{fig:bispectrum}
\end{figure}

As we have just discussed, and Fig.~\ref{fig:TUSR} clearly shows, around the peak scale --- which is the scale of interest for the PBH and SIGW production --- the dominant contribution actually comes from equilateral non-Gaussianity. However, it is also worth noting that the relevant scales for both observables may not exactly coincide with the peak of the power spectrum, due to the strong scale-dependence of the higher order correlations. In fact, it is likely that the bispectrum at different configurations peak at different scales, complicating the prediction for the scale responsible for the PBH and SIGW production. To further illustrate this, in Fig.~\ref{fig:bispectrum} we show the behavior of the squeezed- and equilateral- limit of bispectrum for this model.\footnote{In contrast with what we observed, Ref.~\cite{Ragavendra:2023ret} claims very similar scale-dependence for the squeezed and equilateral non-Gaussianity and the power spectrum. We find this claim rather strange, given the complicated form of the bispectrum. However, a direct comparison is not straightforward since the models considered in Ref.~\cite{Ragavendra:2023ret} are more complicated and the non-Gaussianities are computed numerically.}

The complex scale and shape dependence highlights the importance of using the full correlation functions valid at all scales and for all configurations when inflationary predictions for observables sensitive to non-linearities are concerned. We will study the consequences of these findings  in Sec.~\ref{sec:beta}.

\begin{figure}[h!]
\centering
\includegraphics[width=0.45\textwidth]{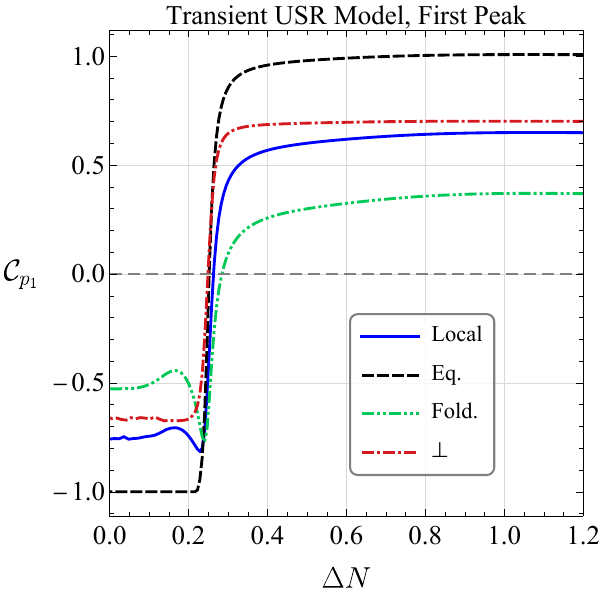}
    \hspace{.5cm}
\includegraphics[width=0.45\textwidth]{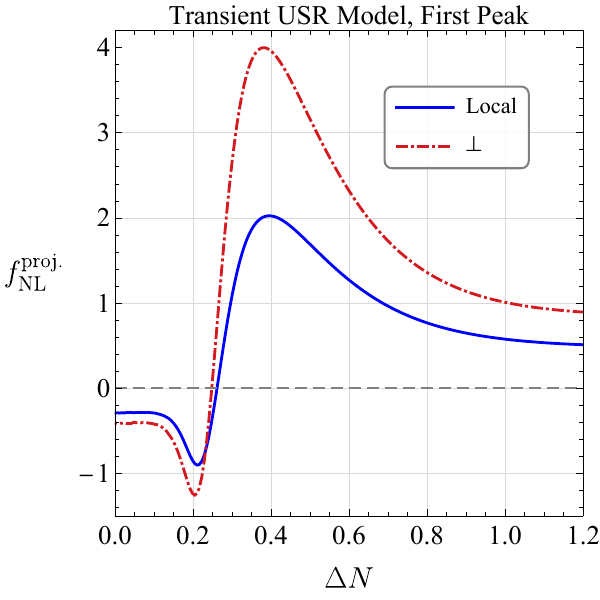}
\caption{
The shape correlators (left) and the projected amplitudes $f_{\mathrm{NL, local}}^{\rm proj.}$ and $f_{\mathrm{NL}, \perp}^{ \rm proj.}$ (right) for the transient USR model as a function of the USR phase duration, $\Delta N$, evaluated at the peak scale of the power spectrum. In this figure, we set $h = -1000$.
}
\label{fig:deltaN-dependence}
\end{figure}

It would also be valuable to examine how the choice of parameters affect the correlation functions. Of particular interest is the duration of the USR phase which also determines the degree of enhancement of the power spectrum. 
Fig.~\ref{fig:deltaN-dependence} shows the shape correlators and the projected amplitudes at the peak scale as a function of $ \Delta N$. 
 As can be seen, the equilateral shape remains dominant over the other templates throughout the range of $\Delta N$ considered. 
 There is also a noticeable transition in Fig.~\ref{fig:deltaN-dependence} from negative to positive correlation around $\Delta N \simeq 0.3$. Interestingly,  this is also the value of $\Delta N$ for which a dip in the power spectrum starts to appear. In other words, looking at different correlations as functions of $\Delta N$, where the dip in the power spectrum appears, the correlation with all non-Gaussian templates at the peak scale becomes small, and the projected amplitudes of non-Gaussianity also drop to zero. Currently, we do not have a  deeper understanding of the connection between these two features and leave further investigations to future works.
 
\begin{figure}[h!]
\centering
\includegraphics[width=0.45\textwidth]{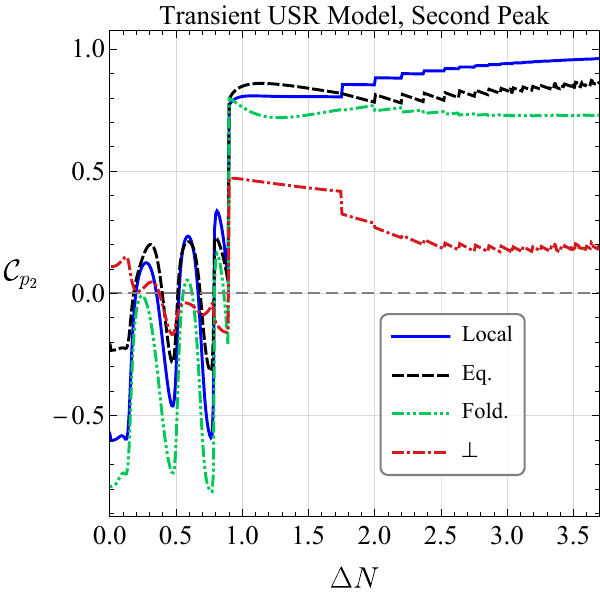}
    \hspace{.5cm}
\includegraphics[width=0.45\textwidth]{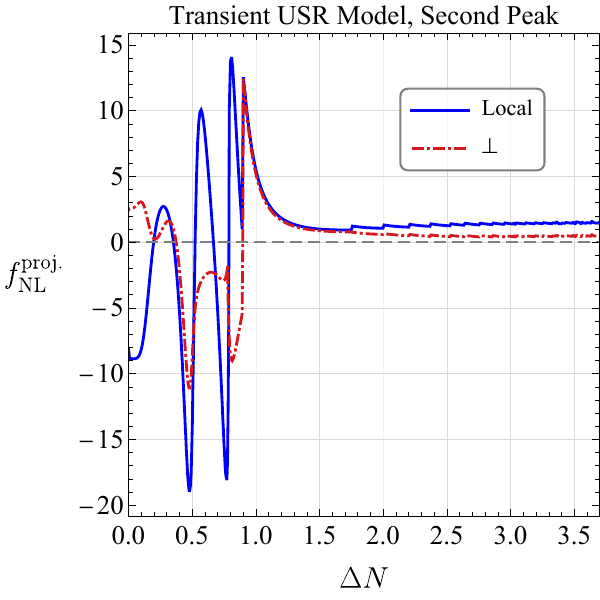}
\caption{
The shape correlators (left) and the projected amplitudes $f_{\mathrm{NL, local}}^{\rm proj.}$ and $f_{\mathrm{NL}, \perp}^{ \rm proj.}$ (right) for the transient USR model as a function of the USR phase duration, $\Delta N$, evaluated at the second-highest peak scale of the power spectrum. The visible breaks in the graphs result from the sudden and discrete shift in the location of the second-highest peak. In this figure, we set $h = -1000$.
}
\label{fig:deltaN-dependence_second}
\end{figure}

 Our previous discussion on the power spectrum (and our findings that will be presented in Sec.~\ref{sec:beta}) motivates the same study for the second-highest peak of the power spectrum, associated with the second phase transition, which we show in Fig.~\ref{fig:deltaN-dependence_second}. Note that as $\tau_e \to \tau_i$, the second phase transition and its associated peak disappear; our transient USR model reduces to the SR-SR model. However, a second-highest peak can still be identified since oscillations in the power spectrum clearly exist, as shown in Fig.~\ref{fig:SR-SR power}. The second-highest peak associated with the second phase transition appears when the two transitions are sufficiently separated; for the current parameters, this occurs for $\Delta N \gtrsim 0.9$, which justifies the change in behavior of ${\cal C}_{p_2}$ near this threshold value. Interestingly, at the second-highest peak, when $\Delta N$ is sufficiently large, the bispectrum is strongly correlated with both local and equilateral shapes. This is very different from what we saw for the highest peak. The observational implications of this finding will be explored in Sec.~\ref{sec:beta}.

Another critical parameter that may affect the shape correlator and projected amplitude at the peaks is $h$. However, $h$ is irrelevant at the highest peak, since the modes around the highest peak are superhorizon at the time of the second transition, where the effect of $h$ appears. Our investigations also suggest that the sensitivity of the second-highest peak to $h$ is also negligible. Therefore, we do not explore this effect further. 

We end this section with a few remarks about the connection between our findings and the general study of the squeezed-limit bispectrum in Ref.~\cite{Namjoo:2024ufv} --- using a generalized consistency relation, first introduced in Ref.~\cite{Namjoo:2023rhq}. In Ref.~\cite{Namjoo:2024ufv}, it is shown that if the long-wavelength mode of the squeezed-limit non-Gaussianity is unaffected by the non-attractor evolution (i.e., if it is frozen before the onset of the non-attractor phase), then the squeezed limit non-Gaussianity vanishes if any peak in the power spectrum is considered as the short-wavelength mode. This agrees with the squeezed-limit of the full bispectrum in the transient USR model, shown in Fig.~\ref{fig:bispectrum}. On the other hand, if the long-wavelength mode does experience the non-attractor evolution, the generalized consistency relation indicates that the size of the squeezed limit non-Gaussianity must be positive. This cannot be observed in Fig.~\ref{fig:bispectrum} because we analyzed a transient non-attractor model and took the extreme long-mode limit $k_\ell \to 0$, which means the long-mode exits the horizon well before the non-attractor phase begins. However, as shown on the right panel of Fig.~\ref{fig:deltaN-dependence}, increasing the duration of the USR phase $\Delta N$, results in a positive local projected size of non-Gaussianity, $f_{\mathrm{NL, local}}^{\rm proj.}$. This aligns with the findings of Ref.~\cite{Namjoo:2024ufv}. For smaller $\Delta N$, fewer long modes exist that are affected by the USR phase and can contribute positively to $f_{\mathrm{NL, local}}^{\rm proj.}$ (and recall that the long modes not influenced by the USR phase do not contribute, according to the consistency relation of Ref.~\cite{Namjoo:2023rhq}). As a result, other, less squeezed correlations dominate in $f_{\mathrm{NL, local}}^{\rm proj.}$, and it appears they tend to produce negative $f_{\mathrm{NL, local}}^{\rm proj.}$.

\section{A simplified estimator}
\label{sec:estimator}
The complexity of the bispectrum from transient non-attractor models motivates consideration of an estimator that is simpler and more economic for next-step studies such as comparing with data or estimating the mass and abundance of PBHs. We have observed that the behavior of the shape correlators for the transient USR model resembles that of the non-BD USR model (Case 1) over a wide range of scales (compare the left panel of Fig.~\ref{fig:7} for $0.1<k|\tau_i|<20$ with the left panel of Fig.~\ref{fig:case1}). Since the latter model has a local-form bispectrum (see Eq.~\eqref{eq:NBD bispectrum}), it suggests the following ansatz, which we refer to as the {\it local-like} (LL) template,
\begin{equation}
	\label{eq:local-like}
	S_{\text{LL}}(k_1,k_2,k_3) \equiv \dfrac{4(k_1k_2k_3)^2}{3(2\pi)^4} \big[P_{k_1} P_{k_2}+ 2 \mathrm{perms.}\big]\, .
\end{equation}
The numerical prefactor --- irrelevant to the shape correlator --- is chosen so that when Eq.~\eqref{eq:shape function} is used, the resulting bispectrum matches that of the local-type non-Gaussianity with $f_{\text{NL}}=1$, assuming it is quantified in real space by ${\mathcal{R}}({\bf x}) = {\mathcal{R}}_G({\bf x}) +\frac35 f_{\text{NL}}{\mathcal{R}}_G^2({\bf x})$, where ${\mathcal{R}}_G$ is a Gaussian random variable.
  
\begin{figure}
\centering
\includegraphics[width=0.45\textwidth]{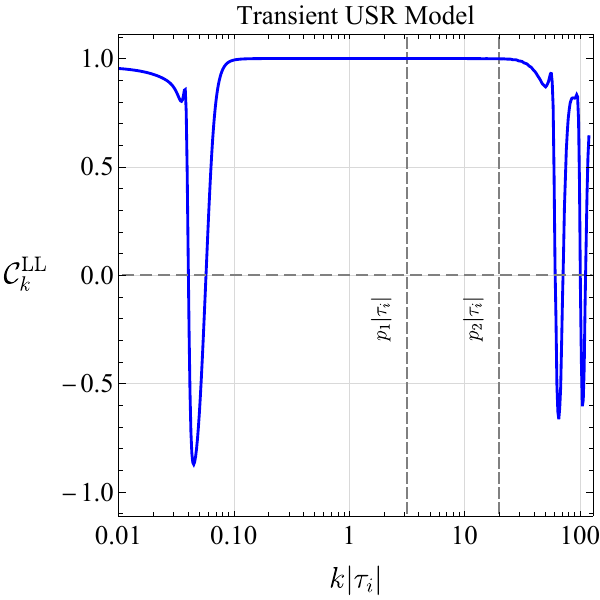}
\hspace{.5cm}
\includegraphics[width=0.45\textwidth]{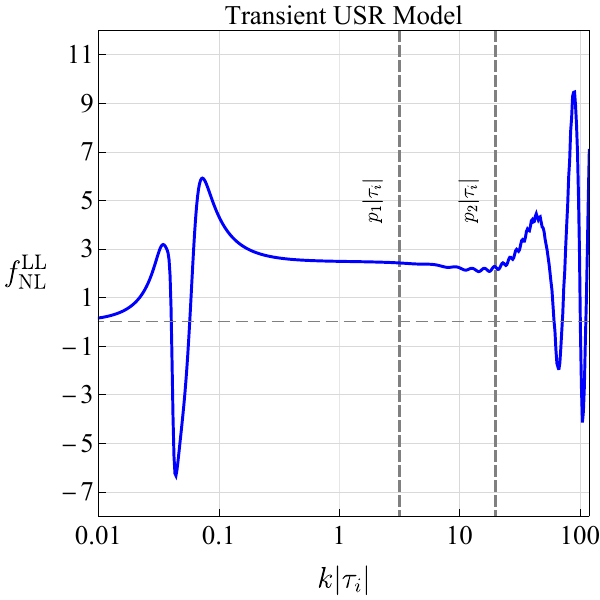}
\caption{
Left: The shape correlator between the local-like template Eq.~\eqref{eq:local-like} and the bispectrum from the transient USR model Eq.~\eqref{eq:transient USR bispectrum}.	Right: The size of local-like non-Gaussianity as implied by the transient USR model, according to Eq.~\eqref{eq:fNLLL}.
}
\label{fig:LL}
\end{figure}

On the left panel of Fig.~\ref{fig:LL}, we show the shape correlator between the bispectrum from the transient USR model and the LL template Eq.~\eqref{eq:local-like}. Interestingly, we observe that the two shapes almost completely overlap across a wide range of scales, including the two highest peaks of the power spectrum. There are notable deviations from a perfect overlap at both small and large scales, but these scales are probably of little relevance for observables due to the smallness of the power spectrum.

The strong overlap between the two shapes motivates further identification of their corresponding bispectra. To do so, we note that in cases of significant overlap, we have
\begin{equation}
\label{eq:equivalence}
S(k_1,k_2,k_3) \approx f_{\text{NL}}^{\text{LL}}(K)\, S_{\text{LL}}(k_1,k_2,k_3)\, ,
\end{equation}
where the symbol $\approx$ indicates that the two sides are equivalent in the sense that their inner product with any other shape gives essentially the same result. In other words, they approximately have the same {\it geometry}. We have also defined $K\equiv \max(k_1,k_2,k_3)$ and $f_{\text{NL}}^{\text{LL}}$ quantifies the size of local-like non-Gaussianity. The peculiar dependence on the momenta is designed so that this factor becomes irrelevant when scale-dependent shape correlators are computed (it drops out from Eq.~\eqref{eq:shape-correlator}, since $K=k$ when the bispectrum is expressed in terms of $k$, $x$ and $y$, according to Eq.~\eqref{eq:x and y}). However, note that it does appear in the inner product which allows us to use Eq.~\eqref{eq:equivalence} to identify\footnote{This definition resembles ``fudge factor" defined in Ref.~\cite{Babich_2004}.}
\begin{equation}
	\label{eq:fNLLL}
	f_{\text{NL}}^{\text{LL}}(k) \to  \dfrac{\langle S , S_{\text{LL}}  \rangle_k}{\langle S_{\text{LL}} , S_{\text{LL}}  \rangle_k}.
\end{equation}
Note that this identification assumes complete overlap; caution is needed when interpreting the result for scales where this assumption does not hold. Additionally, this differs from Eq.~\eqref{eq:fNL} mainly because $S_{\text{LL}}$ also includes the power spectrum. Since we have defined several non-Gaussianity amplitudes (and will introduce yet another one in the next section), we summarize them in Table \ref{tab:fNLs} to prevent confusion.

\begin{table}[t]
\centering
\caption{Different definitions of the size of non-Gaussianity discussed in this paper.}
\renewcommand{\arraystretch}{2} 
\begin{tabular}{|c|c|}

\hline
\textbf{Quantity} & \textbf{Definition} \\
\hline

Size of local non-Gaussianity
& via $\mathcal{R}(\mathbf{x})=\mathcal{R}_G(\mathbf{x})+\frac{3}{5} f_{\mathrm{NL}} \mathcal{R}_G^2(\mathbf{x})$ \\

Projected size of non-Gaussianity, Eq.~\eqref{eq:fNL}
& $f_{{\rm NL}, T}^{\rm proj}(k) = 
\dfrac{\langle S , S_T \rangle_k}{\langle S_T , S_T \rangle_k \mathcal{P}_k^2}$ \\

Local-like size of non-Gaussianity, Eq.~\eqref{eq:fNLLL}
& $f_{\rm NL}^{\rm LL}(k) = 
\dfrac{\langle S , S_{\rm LL} \rangle_k}{\langle S_{\rm LL} , S_{\rm LL} \rangle_k}$ \\

Effective size of non-Gaussianity, Eq.~\eqref{eq:fNL effective}
& $f_{\rm NL}^{\rm eff}(k) = \dfrac{\mathcal{S}_3}{\sigma^4}$ \\
\hline
\end{tabular}
\label{tab:fNLs}
\end{table}

Using this relation along with Eqs.~\eqref{eq:shape function} and \eqref{eq:equivalence}, one can construct an estimator for the bispectrum as follows
\begin{equation}
	\label{eq:B local-like}
	B_{\mathcal{R}}^{\text{LL}}(k_1,k_2,k_3) \equiv \dfrac{6}{5} f_{\text{NL}}^{\text{LL}}(K) \big[P_{k_1} P_{k_2}+ 2 \mathrm{perms.}\big]\, .
\end{equation}

On the right panel of Fig.~\ref{fig:LL}, we show the behavior of $f_{\text{NL}}^{\text{LL}}(k)$. We see that there is a running in $f_{\text{NL}}^{\text{LL}}$ even within the range of scales where the overlap between the two shapes is nearly complete. Thus, the required local-like bispectrum is {\it not} equivalent to the local bispectrum, for which  $f_{\text{NL}}^{\text{LL}}$ is a constant. Nonetheless there is also a plateau in a range of scales including the highest peak (but not the second-highest peak which, as we will discuss in Sec.~\ref{sec:beta}, is also of interest).\footnote{It is interesting to note that a similar plateau is observed for the size of non-Gaussianity in equilateral configurations in Ref.~\cite{Davies:2021loj}, where various scenarios with multiple phases of constant $\eta$ are studied. We also note that the ``plateau" actually has a small negative slope. Very recently, Ref.~\cite{Escriva:2025ftp} appeared that effectively adds a scale-dependent effect to the scale-invariant $\delta N$ formula for $\cal R$. In its perturbative form, their results predict a negative running of the size of non-Gaussianity which seems to align with our findings. We thank J. Garriga for pointing this out to us. A more detailed comparison would be worthwhile which we leave for the future.}

It is worth emphasising that the estimator Eq.~\eqref{eq:B local-like} may only be useful {\it a posteriori}, meaning that one must first calculate the full bispectrum, from which $f_{\text{NL}}^{\text{LL}}$ can be derived. Afterwards, the local-like bispectrum is fully determined and can be used for other purposes, such as comparing with data or investigating the PBH and SIGW formation as predicted by these models.  

We have confirmed that the same estimator applies to the SR-SR  model, while its applicability to the non-BD USR model is obvious. We have also confirmed that the overlap over the scale range of interest persists regardless of the choice of parameters, particularly $h$ and $\Delta N$. The model examined here can be made more complex by adding a sharp step to the potential immediately after the USR phase \cite{Namjoo:2024ufv,Cai_2022}. For simplicity, we have not included the resulting bispectrum in this paper. Nonetheless, we have verified that the same estimator (albeit with a different $f_{\text{NL}}^{\text{LL}}$) remains a good approximation to the full bispectrum across a broad range of scales. This indicates that the LL estimator could have a much wider applicability and might stay valid for a broader class of transient non-attractor inflationary models.\footnote{It is worth noting that Ref.~\cite{Taoso:2021uvl} analyzed a different transient non-attractor model and concluded that a local-type bispectrum suffices to describe the leading three-point correlation function. We find the reason for their differing claim to be that they completely ignore the bispectrum from the transient non-attractor phase. Instead, they obtain their leading bispectrum solely from a late-time field redefinition, which is clearly of the local-type. The reason for the dominance of this contribution is that they consider a large value of $\eta$ for the final stage of inflation. We find this assumption unrealistic since the non-attractor phase is expected to be followed by a slow-roll, attractor inflation, during which the slow-roll parameters are small (although a short period of relaxation can occur, as described by our model).}
 We will defer further investigations into this to future work. 
 
\begin{figure}
	\centering
	\includegraphics[width=0.45\textwidth]{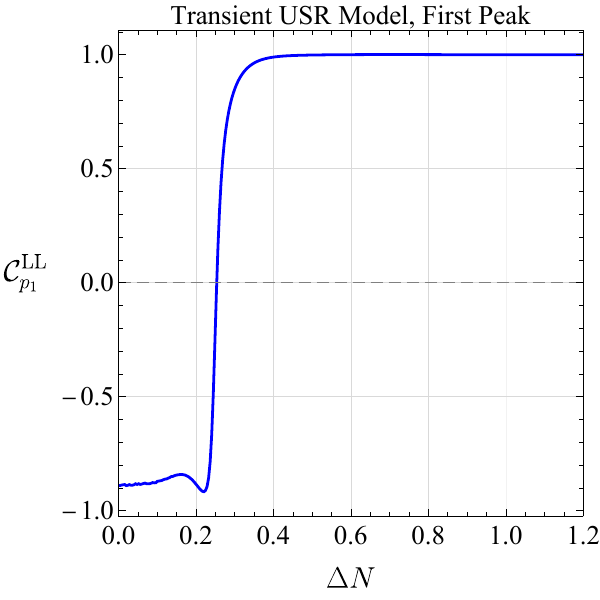}
	\hspace{.5cm}
	\includegraphics[width=0.45\textwidth]{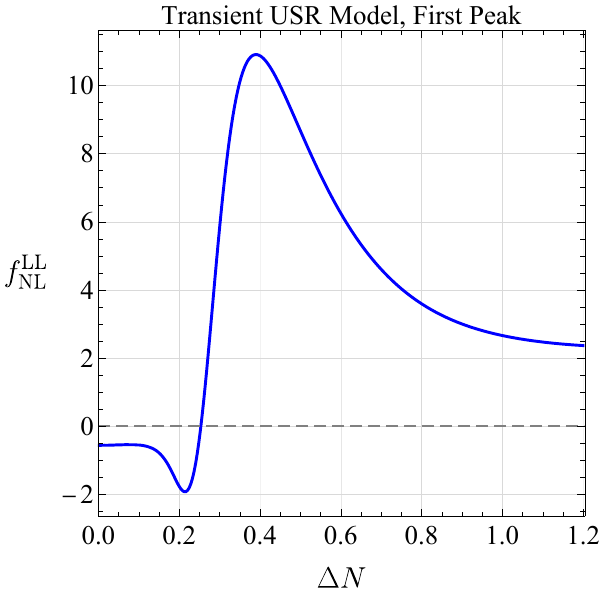}
	\caption{
		 The shape correlator between the local-like template Eq.~\eqref{eq:local-like} and the shape from the transient USR model as a function of $\Delta N$, the duration of the USR phase (left), and the corresponding size of local-like non-Gaussianity (right).
	}
	\label{fig:LLDeltaN}
\end{figure}

As a final remark, it is worth noting that while the validity of the estimator appears to be robust against the details of the model, $f_{\text{NL}}^{\text{LL}}$ does vary and, in some cases, varies significantly. As an example, in Fig.~\ref{fig:LLDeltaN}, we illustrate the behaviour of the shape correlator and  $f_{\text{NL}}^{\text{LL}}$ at the peak scale of the power spectrum as functions of $\Delta N$, the duration of the USR phase. We observe that while the correlation remains nearly complete across the entire range,  $f_{\text{NL}}^{\text{LL}}$ varies considerably with $\Delta N$. We also see that for sufficiently large $\Delta N$, $f_{\text{NL}}^{\text{LL}}$ approaches 2.5, the size of local non-Gaussianity from pure USR inflation. This may roughly be understood, though not entirely justified,  by the fact that for large $\Delta N$, the peak scale leaves the horizon during the USR phase, which is not the case if  $\Delta N$ is small. This observation, together with the plateau in Fig.~\ref{fig:LL}, suggests that the $\delta N$ formalism --- despite being unable to consistently capture scale-dependencies --- may not be far off after all. This deserves further investigation, which we leave for the future.

\section{Implications for Cosmological Observables}
\label{sec:beta}
\begin{figure}[t]
    \centering
    \includegraphics[width=0.45\textwidth]{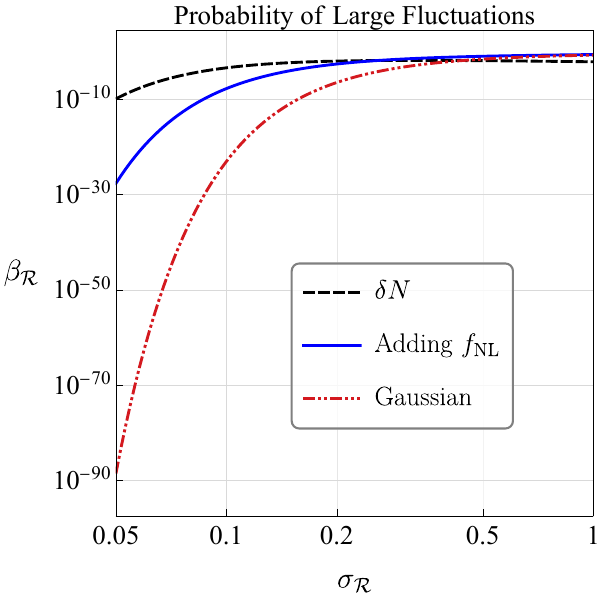}
    \caption{
The probability of  larger-than-one curvature perturbation $\beta_\mathcal{R}$ for the pure USR model evaluated via, Gaussian PDF and the modified PDF using the perturbative $\delta N$ formula up to the second-order (i.e., including the term proportional to $f_\mathrm{NL}$), and  the non-perturbative $\delta N$ formula. 
    }
    \label{fig:GfdN}
\end{figure}

The results of previous sections, suggest that considering the bispectrum at all configurations is needed to make precise assessment of the observational effects. 
To further explore this issue, we calculate the probability of large fluctuations, which is expected to be a relevant study to the abundance of PBHs. However, note that for a complete result, one needs access to the full probability density function (PDF), especially at the tail, where a non-perturbative analysis is required \cite{Hooshangi:2021ubn}. Such a comprehensive study is beyond the scope of this paper and, to our knowledge, a suitable tool for this non-perturbative analysis, applicable to the considered models, has not yet been developed. Therefore, we can only study how the perturbative calculations presented here influence the leading-order, Gaussian predictions. Depending on the parameter choices, this may serve as a good indicator of what a full analysis might yield. To see this, it is useful to compare the perturbative analysis with a non-perturative result for a simple well-known case.  In Fig.~\ref{fig:GfdN} we show the probability of realizing larger-than-one curvature perturbation $\beta_{\cal R}$ as a function of $\sigma_\mathcal{R} \equiv {\cal P}_{\cal R}^{1/2}$ in the pure USR model, where the USR inflation is assumed to be the only relevant physics that the modes of interest experience before they freeze and stop evolving \cite{Namjoo_2012}. Assuming that the inflaton's fluctuations are purely Gaussian, the perturbative and non-perturbative PDF for $\cal R$ can be obtained using the $\delta N$ relation \cite{Sasaki_1996, Sasaki_1998, Wands_2000, Lyth_2005}. This relation for the pure USR model is well-known and can be seen e.g., in Refs.~\cite{Chen:2013eea,Biagetti:2021eep,Hooshangi:2023kss}, so we do not repeat them here. As can be seen in Fig.~\ref{fig:GfdN}, taking the local bispectrum $f_\mathrm{NL}=2.5$ into account brings $\beta_\mathcal{R}$ much closer to its non-perturbative value, although it may still be off by several orders of magnitude, depending on the value of $\sigma_{\cal R}$. We conclude that, while the full picture remains to be studied, the inclusion of bispectrum is a significant improvement and we proceed by only considering this effect. 

It is worth noting that the relevant quantity for the PBH formation is the smoothed density contrast during the radiation-dominated era, rather than the curvature perturbation \cite{Ezquiaga_2020, Hooshangi:2021ubn, Biagetti_2018}. Since we know the statistics of the curvature perturbation up to the second-order in perturbation theory, we relate these two quantities at that order. As we will see, this connection introduces another scale-dependence into the analysis. We then provide an effective description of the distribution of smoothed density perturbations. Then, we calculate the probability of large smoothed density contrast for the transient USR model when second-order perturbations are included.

%\subsection{From curvature perturbation to smoothed matter density fluctuation}
%\label{sec:From PCP to MDF}
The density contrast in the real space $ \delta$, in the radiation era and at large scales is related to the curvature perturbation
% $\left(\mathcal{R} \right)$
via \cite{Shibata_1999, Harada_2015}, 
\begin{equation}
\label{eq:delta to R}
\delta =- \frac{8}{9} \left(\frac{1}{a H} \right)^2 e^{-2 \mathcal{R}} \left( e^{- \mathcal{R}/2} \nabla^2 e^{\mathcal{R}/2} \right).
\end{equation}
Expanding Eq.~\eqref{eq:delta to R} up to $\mathcal{O}(\mathcal{R}^2)$ yields,
\begin{equation}
\label{eq:delta to R expanded}
\delta\simeq  -\frac{4}{9} \left( \frac{1}{\tilde a H} \right)^2  \big(  \nabla^2 \mathcal{R} + \frac{1}{2} \nabla \mathcal{R} \cdot \nabla \mathcal{R} \big) \, ,
%\equiv \delta_l + \delta_{nl},
\end{equation}
where $\tilde a =a \, e^{\mathcal{R}}$ is the effective scale factor in the presence of the curvature perturbation.
%where we have defined $\delta_l \equiv -4/9 (a e^{\mathcal{R}} H)^{-2} \nabla^2 \mathcal{R}$ and $\delta_{nl} \equiv -4/9 (a e^{\mathcal{R}} H)^{-2} \nabla \mathcal{R} \cdot \nabla \mathcal{R}$ for later convenience.
In Fourier space, this relation takes the following form 
\begin{equation}
\label{eq:delta in Fourier space}
\delta_k = \lambda_c^2 \, k^2 \mathcal{R}_k +  \frac{ \lambda_c^2}{2} \int \frac{\mathrm{d}^3 \textbf{q}}{\left( 2 \pi \right)^3} \left( \textbf{q} \cdot \left(\textbf{k} - \textbf{q} \right) \right) \mathcal{R}_{k-q} \mathcal{R}_q\, ,
%\delta_{l, k} = k^2 \mathcal{R}_k, \quad \delta_{nl, k} = \frac{1}{2} \int \frac{\mathrm{d}^3 \textbf{q}}{\left( 2 \pi \right)^3} \left( \textbf{q} \cdot \left(\textbf{k} - \textbf{q} \right) \right) \mathcal{R}_{k-q} \mathcal{R}_q,
\end{equation}
where we have defined $\lambda_c \equiv \frac{2}{3 \tilde a H}$ which is proportional to the comoving horizon length. 

The quantity of interest is  the smoothed matter density contrast defined by
\begin{equation}
\label{eq:smoothing}
\tilde{\delta}(x) = \int \delta(x^{\prime}) \bar W_R(x -x^{\prime}) \mathrm{d}^3 \mathbf{x^{\prime}} =  \int \frac{\mathrm{d}^3 \textbf{k}}{(2 \pi)^3} \: \delta_k \: W_R(k) e^{i \textbf{k}.\textbf{x}}\, ,
\end{equation}
where $\bar W_R(x - x^{\prime})$ and $W_R(k)$ are the window function and its Fourier transform, respectively, and play the role of smoothing fluctuations at the comoving scale $R$.
Throughout this paper, we consider a top-hat window function in Fourier space and set aside complications associated with the choice of window function \cite{Young_2019},
\begin{equation}
\label{eq:Top-hat window function}
W_R (k) = \theta(1-k R).
\end{equation}
 
 The  variance of the smoothed matter density contrast can be calculated using Eqs.~\eqref{eq:delta to R expanded} and~\eqref{eq:smoothing} which yields
 \begin{equation}
 \label{eq:Smoothed Variance}
 \begin{aligned}
 \sigma^2  = \langle\tilde{ \delta}^2 \rangle = &\lambda_c^4 \int_0^\infty W_R (k)^2 k^4 \mathcal{P}_k \mathrm{d}\log k 
 \\
& -\lambda_c^4 \int \frac{\mathrm{d}^3 \textbf{k}_1}{(2 \pi)^3} \int \frac{\mathrm{d}^3 \textbf{k}_2}{(2 \pi)^3}  W_R(k_1)^2 \: k_{1}^2 \left( \textbf{k}_2 \cdot \textbf{k}_{12} \right) B_{\mathcal{R}}(k_1 , k_2 , k_{12}),
 \end{aligned}
 \end{equation}
 where
 % $\mathcal{P}_{\mathcal{R}} (k) = \frac{k^3}{2 \pi^2} P_{\mathcal{R}} $  and
 $\textbf{k}_{12} = \textbf{k}_1 + \textbf{k}_2$.
 % We have also used the definition of the bispectrum $B_{\mathcal{R}}(k_1 , k_2 , k_3)$,
%\begin{equation}
%\label{eq:Bispectrum}
%\langle \mathcal{R}_{\textbf{k}_1}  \mathcal{R}_{\textbf{k}_2}  \mathcal{R}_{\textbf{k}_3} \rangle \equiv (2 \pi)^3 \delta^3(\textbf{k}_1 + \textbf{k}_2 + \textbf{k}_3 ) B_{\mathcal{R}}(k_1 , k_2 , k_3).
%\end{equation} 
 Note that the second term in Eq. \(\eqref{eq:Smoothed Variance}\) is subdominant and, from now on, will be neglected.

The three-point correlation function of $\tilde \delta$ may be expressed by
%\begin{equation}
%\label{eq:delta3}
%\delta^3 = \langle \delta_R \rangle = \langle \delta_l^3 \rangle + 3 \langle \delta_l^2 \delta_{nl} \rangle + \mathcal{O} (\mathcal{R}^5).
%\end{equation}
%Again, performing smoothing in the Fourier space Eq.~\eqref{eq:delta3} leads to,
\begin{equation}
\label{eq:delta3 in Fourier}
\begin{aligned}
\mathcal{S}_3 \equiv \langle \tilde \delta^3\rangle=  \mathcal{S}_{3,l} + \mathcal{S}_{3,nl}\, ,
\end{aligned}
\end{equation}
where 
\begin{equation}
\begin{aligned}
&\mathcal{S}_{3,l} = 3 \lambda_c^6 \int \frac{\mathrm{d}^3 \textbf{k}_1}{(2 \pi)^3} \int \frac{\mathrm{d}^3 \textbf{k}_2}{(2 \pi)^3}  W_R(k_1) W_R(k_2) W_R(k_{12}) \: k_1^2 \:k_2^2 \left( \textbf{k}_1 \cdot \textbf{k}_2 \right) P_{k_1} P_{k_2},\\
&\mathcal{S}_{3,nl} = \lambda_c^6 \int \frac{\mathrm{d}^3 \textbf{k}_1}{(2 \pi)^3} \int \frac{\mathrm{d}^3 \textbf{k}_2}{(2 \pi)^3}  W_R(k_1) W_R(k_2) W_R(k_{12}) \: k_1^2 \:k_2^2 \, k_{12}^2 \, B_{\mathcal{R}}(k_1 , k_2 , k_{12}).
\end{aligned}
\end{equation}
$\mathcal{S}_{3,nl}$ originates from the  three-point function of the primordial fluctuations while $\mathcal{S}_{3,l}$ is induced from late time non-linearities --- which exists even if the primordial curvature perturbation is purely Gaussian.  For $f_{\text{NL}} \sim {\cal{O}}(1)$,  which is of interest here, the two contributions will be comparable. 

To account for the effect of non-linearities on the PDF of $\tilde \delta$, we first compute its skewness  which is defined  by \cite{LoVerde:2007ri,Matsubara:2022nbr,Shandera:2012ke}
\begin{equation}
\label{eq:Skewness}
\gamma \equiv  \frac{\mathcal{S}_3}{\sigma^3}\, . % = \gamma_l + \gamma_{nl}\, ,
\end{equation}
Then, we consider the following ansatz:
\begin{equation}
\label{eq:delta tylor}
\tilde{\delta}_{\mathrm{eff}} = \tilde{\delta}_G + f_{\mathrm{NL}}^{\mathrm{eff}}\, \tilde{\delta}_{G}^{ 2}\, ,
\end{equation}
where $ \tilde{\delta}_G$ is a Gaussian random variable with the same variance as in Eq.~\eqref{eq:Smoothed Variance}. We determine $f_{\mathrm{NL}}^{\mathrm{eff}}$ by demanding  the above expression to yield the same skewness as obtained in  Eq.~\eqref{eq:Skewness}:
\begin{equation}
\label{eq:fNL effective}
f_{\mathrm{NL}}^{\mathrm{eff}} = \frac{ \gamma}{ \sigma}= \frac{ \mathcal{S}_3}{ \sigma^4} .
\end{equation}
It should be noted that Eq.~\eqref{eq:delta tylor} effectively accounts for the contribution of the primordial non-linear effects up to the bispectrum at all configurations to the {\it local} non-linearities of the density fluctuations. Primordial non-linear effects also induce non-local imprints on the matter fluctuations, which the above definition does not capture. However, such non-local contributions primarily affect clustering, which we do not study in this paper. For our purpose — namely, estimating the probability of large fluctuations — the statistics of local fluctuations suffices.

From Eq.~\eqref{eq:delta tylor}, the effective PDF for $\tilde \delta$ is given by \cite{Byrnes_2012},
\begin{equation}
\label{eq:Distribution Function}
\begin{aligned}
&P_{\mathrm{NG}}(\tilde{\delta}) \mathrm{d} \tilde{\delta}=\frac{\mathrm{d} \tilde{\delta}}{\sqrt{2 \pi} \sigma \sqrt{1+4 f_{\mathrm{NL}}^{\mathrm{eff}}\big( f_{\mathrm{NL}}^{\mathrm{eff}} \sigma^2+\tilde{\delta} \big)}}\left(\varepsilon_{-}+\varepsilon_{-}\right),\\
\end{aligned}
\end{equation}
where
\begin{equation}
\varepsilon_{ \pm}=\exp \left[-\frac{1}{2}\bigg(\frac{h_{ \pm}^{-1}(\tilde{\delta})}{\sigma}\bigg)^2\right], \quad h_{ \pm}^{-1}(\delta_R)=\frac{1}{2 f_{\mathrm{NL}}^{\mathrm{eff}}}\left[-1 \pm \sqrt{1+4 f_{\mathrm{NL}}^{\mathrm{eff}}\big(f_{\mathrm{NL}}^{\mathrm{eff}} \sigma^2+\tilde{\delta} \big)}\right].
\end{equation}

As a crude estimate of the effect of non-Gaussianity on the PBH production, we  calculate the probability of large fluctuations as follows
\begin{equation}
\label{eq:beta}
	\beta = \int_{\delta_c}^{\delta_{\mathrm{max}}} P_{\mathrm{NG}}(\tilde{\delta}) \, \mathrm{d} \tilde{\delta},
\end{equation}
where $\delta_c$ is a threshold, required for the PBH production, and  $\delta_{\mathrm{max}}$ is the maximum possible fluctuation that should be determined according to the sign of $f_{\mathrm{NL}}^{\mathrm{eff}}$; for $f_{\mathrm{NL}}^{\mathrm{eff}} > 0$, $\delta_{\mathrm{max}} \to \infty$ while for $f_{\mathrm{NL}}^{\mathrm{eff}}< 0$, $\delta_{\mathrm{max}}$ is given by
\begin{equation}
	\label{eq:delta max}
	\delta_{\mathrm{max} }=-\frac{1}{4 f_{\mathrm{NL}}^{\mathrm{eff}}}\left(1+4 f_{\mathrm{NL}}^{\mathrm{eff}\, 2}\, \sigma^2\right).
\end{equation}
The actual value of the threshold is influenced by several factors, including the choice of window function and the collapse time.\footnote{It is demonstrated in Ref. \cite{Kehagias_2019} that the threshold may also depend on the amplitude of  non-Gaussianity. The study of Ref. \cite{Kehagias_2019} is limited to local non-Gaussianity. Nonetheless, naively employing those results to our scenario, we find that this effect does not have a significant impact on our main results. Therefore, we will set aside this subtlety in this paper.}
Based on the simulations and numerical calculations, it appears that when using the top-hat window function, it is appropriate to take \cite{Young_2019}
\begin{equation}
\label{eq:Threshold}
\delta_c \simeq 0.59.
\end{equation}

\begin{figure}[t]
	\centering
	\includegraphics[width=0.45\textwidth]{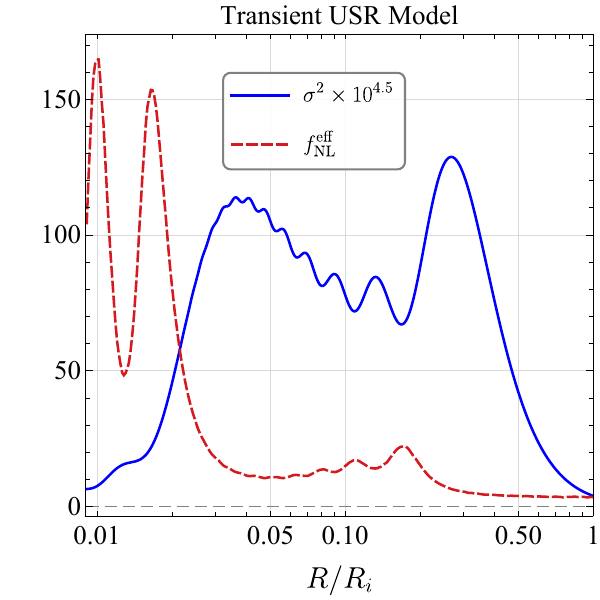}
	\caption{The smoothed variance of density contrast and the effective amplitude of non-Gaussianity as functions of comoving scale for the transient USR model. $R_i$ is the comoving horizon at the first transition, i.e. $\tau_i$. The parameters are the same as in Fig.~\ref{fig:power-spectrum}.}
	\label{fig:7}
\end{figure}

With these ingredients, we are prepared to analyze the probability of large fluctuations in the transient USR model.\footnote{Note that, for the calculations in this section, it does not matter much whether the full bispectrum or the local-like estimator is used, since --- as we have shown in Sec.~\ref{sec:estimator} --- the two bispectra share the same leading features at the scales of interest. This has also been explicitly confirmed by direct calculations which only shows some difference near the highest peak of $\beta$, corresponding to the second-highest peak of the power spectrum, where small deviations from a complete overlap is visible in Fig.~\ref{fig:LL}. It is also tempting to use a local-type non-Gaussianity, corresponding to the local-like non-Gaussianity with a constant $f_{\text{NL}}^{LL}$ that can be fixed to the value on the plateau shown in the left panel of Fig.~\ref{fig:LL}. We find that, although this simplification changes $\beta$ near the highest peak of the power spectrum only by a factor of a few,
%(somewhat consistent with the claims of Refs.~\cite{Davies:2021loj})
 near the second-highest peak they differ by about a factor of 10. Therefore, a purely local-type non-Gaussianity does not accurately capture the predictions of these models. In any case, even if one ignores this difference, the actual value of $f_{\text{NL}}$ still requires the full calculation of the bispectrum.} In Fig.~\ref{fig:7}, we display the effective amplitude of non-Gaussianity and the variance of the smoothed density contrast as functions of comoving scale $R$ for the transient USR model. The two primary peaks in $\sigma^2$ are linked to the two highest peaks in the power spectrum; see Fig.~\ref{fig:power-spectrum}. Interestingly, we observe that $f_{\mathrm{NL}}^{\mathrm{eff}}$ is larger near the second-highest peak, suggesting that non-linearities are more significant there than around the highest peak. Overall, the distinct scale-dependence of $\sigma$ and $f_{\mathrm{NL}}^{\mathrm{eff}}$ indicates a possible trade-off between the size of the variance and the importance of non-linearities. Combining these effects may result in a significantly wider range of scales where the PBH formation can occur efficiently. Clearly, the dominant scale for the PBH formation depends on the specifics of the model. We will investigate it for a particular set of parameters within the transient USR model. 

\begin{figure}[t]
    \centering
    \includegraphics[width=0.45\textwidth]{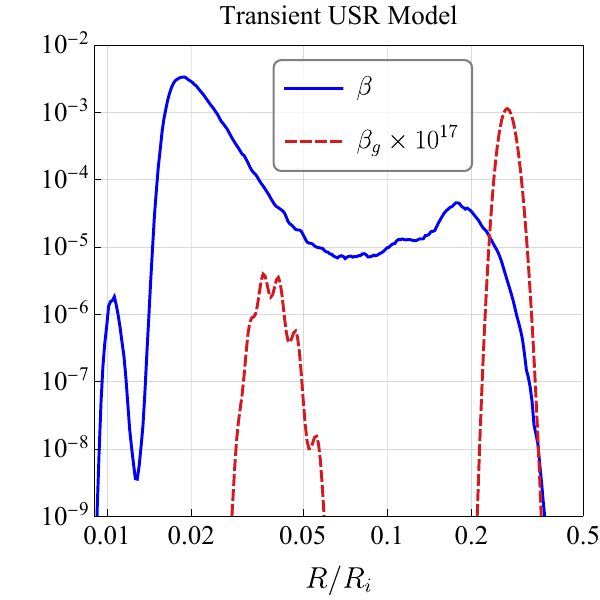}
 %   \hspace{.4cm}
%    \includegraphics[scale = 0.5]{Figures/beta extended.pdf}
    \caption{The probability of large density fluctuation for the transient USR model as a function of comoving scale. The parameters are the same as in Fig.~\ref{fig:power-spectrum}. $R_i$ is the size of the horizon at the time of the first transition. }
    \label{fig:beta}
\end{figure}

Fig.~\ref{fig:beta} shows how $\beta$ varies with the comoving scale $R$. For comparison, we also display the behavior of $\beta$ when all non-linearities are ignored. It can be seen clearly that including non-Gaussianities significantly alters both the value of $\beta$ and its dependence on scale. Notably, there is a much broader range of scales where efficient PBH formation (corresponding to large $\beta$) can occur when non-Gaussianities are included. The peak of $\beta$ appears near the second-highest peak of the variance, which differs in comoving scale from the highest peak by a factor of 10. This aligns with --- but is slightly larger than --- the separation of the two main peaks in the primordial power spectrum (see Fig.~\ref{fig:power-spectrum}). Consequently, non-linearities not only influence the abundance of PBHs but also significantly impact their mass range.\footnote{Note that a change in comoving scale by a factor of 10 corresponds to a PBH mass change of a factor of 100.} This underscores the importance of moving beyond the Gaussian approximation and considering the full bispectrum when connecting inflationary dynamics to the PBH formation. It would be valuable to investigate similar effects on the SIGWs, which we reserve for future research.

\section{Conclusion}
\label{sec:conclusion}
In this work, we calculated the full bispectrum for inflationary models with transient non-attractor phases and studied its {\it geometry}, i.e, its dependence on scale and shape. We found that the bispectrum generally exhibits a complex {\it geometry} due to different physics governing different scales. In transient non-attractor models, non-attractor evolution, phase transitions, and particle production coexist, each influencing the evolution of different modes and their correlations.
As general rules, the bispectrum at scales affected by sharp transitions and particle production (typically including the peak scale of the power spectrum) is strongest near the equilateral configuration, while non-attractor phases tend to produce correlations near the squeezed configuration.  In contrast, despite deviations from the Bunch-Davies vacuum, it is unlikely for these scenarios to create a strong correlation near the folded configuration, since positive and negative frequency mixing is not efficiently produced.

Additionally, we have shown that the full bispectrum can be approximated by a local-like estimator that looks like the bispectrum from local-type non-Gaussianity, but with the twist that the size of non-Gaussianity (which is a constant in the local non-Gaussianity) varies as a function of scale. We have observed that this simplification remains valid across a broad range of scales and parameter space, and for different inflationary models involving transient non-attractor phases. Nevertheless, this finding should not be overstated, as fully identifying the local-like bispectrum --- particularly its scale-dependent size --- requires knowledge of the full bispectrum, which is model-dependent. Further investigation into the scope of this estimator's generality and applicability is an interesting direction for future research. 

Our findings suggest that a thorough analysis of non-linearities, including correlations across all configurations, is necessary for accurately predicting the PBH mass and abundance as well as the frequency and amplitude of SIGWs. 
As a more explicit consequence of our results, we estimated the probability of large smoothed density contrast, considering  up to the second-order effects in perturbation theory. Our results suggest that because the scale-dependence differs significantly between the power spectrum and bispectrum, accounting for the latter tends to shift or widen the mass window of PBHs;  or open up a new one, near the second-highest peak of the power spectrum, provided it is high enough. To our knowledge, this effect has been overlooked in the literature. A similar analysis for the effect of the full bispectrum on the frequency and amplitude SIGWs is worthwhile, which we leave for future work.

Finally, we emphasize that our estimation oversimplifies the relationship between PBH formation and non-linearities in various ways, and our results should be viewed as illustrative rather than precise predictions of the models considered. In particular, we have focused solely on the bispectrum. Higher-order non-linear effects beyond the three-point function may display different scale-dependences and leave additional imprints, and, at least for the PBHs, need to be considered. Studying these higher-order contributions remains an important direction, and our current study is only the first step, also motivating the development of new non-perturbative techniques for studying scenarios with strongly scale- and configuration-dependent correlations.

\acknowledgments
We thank H. Firouzjahi, J. Garriga, M. Noorbala, A. Riotto and  V. Vennin for fruitful discussions.

\appendix

\section{Full-Shape Bispectrum of the Transient USR Model}
\label{sec:details}
In this appendix, we present the full technical steps involved in computing the bispectrum for the transient USR model. A  particular limit of the results correspond to the SR-SR model of Sec.~\ref{sec:SR-SR}, which will also be occasionally  discussed.

% We begin by reviewing the background evolution. Then we briefly review the evolution of the mode functions in each phase of inflation, i.e., SR, USR, and SR.  Finally, we compute the bispectrum using the in-in formalism.

\subsection{Background Evolution}
\label{sec:background}
The dynamics of single scalar field inflation is described by the following equations of motion,
\begin{equation}
\label{eq:KG}
3 M_{\rm Pl}^2 H^2 = \frac{1}{2} \dot{\phi}^2 + V(\phi),\quad \ddot{\phi} + 3 H \dot{\phi} + V'(\phi)= 0,
\end{equation}
where $H \equiv \dot{a}/a$ is the Hubble parameter, $a(t)$ is the scale factor, and over dots denote derivatives with respect to cosmic time.
Given the inflaton potential in Eq.~\eqref{eq:Potential}, the evolution of the scalar field and the background expansion can be fully determined.
For our purpose, however, we only need the first and second slow-roll (SR) parameters, which are defined as,
\begin{equation}
\label{eq:Epsilon and Eta}
\epsilon \equiv -\frac{\dot{H}}{H^2},\quad \eta \equiv \frac{\dot{\epsilon}}{H \epsilon}.
\end{equation}

In the transient USR model which is described by the potential Eq.~\eqref{eq:Potential}, the initial phase of inflation is a slow-roll phase, where the inflaton field follows the attractor trajectory. Therefore, in this phase the SR parameters remain approximately constant,
\begin{equation}
\label{eq:SR-phase}
\text{For} \: \tau < \tau_i: \quad \epsilon \simeq \epsilon_i, \quad \eta \simeq 0. 
\end{equation}
The USR phase occurs on the flat segment of the potential, $\phi_e < \phi < \phi_i$, where  the inflaton field speed down exponentially and the SR parameters are given by
\begin{equation}
\label{eq:USR phase}
\text{For} \: \tau_i < \tau < \tau_e: \quad \epsilon(\tau) = \big( \frac{\tau}{\tau_i} \big)^6 \epsilon_i, \quad \eta = -6.
\end{equation}

According to Eqs.~\eqref{eq:SR-phase} and \eqref{eq:USR phase}, the SR parameters at the first transition are given by,
\begin{equation}
\label{eq:First Transition}
\text{For} \: \tau_i^{-} \leq \tau \leq \tau_i^{+}: \quad \epsilon = \epsilon_i,\quad \eta = -6\, \theta(\tau - \tau_i)
\end{equation}
After the USR phase ends at $\phi = \phi_e$, the inflaton field exits the flat segment of the potential and gradually relaxes to the second SR attractor. The evolution of the SR parameters during this relaxation can be obtained solving Eq.~\eqref{eq:KG}, 
\begin{equation}
\label{eq:Relaxation}
\text{For} \: \tau_e < \tau: \quad \epsilon(\tau) = \epsilon_e \bigg( \frac{h}{6} -  \big(1 + \frac{h}{6} \big) \big(\frac{\tau}{\tau_e} \big)^3 \bigg)^2, \quad \eta = -\frac{6 (6 +h)}{(6+h) - h \big( \frac{\tau_e}{\tau} \big)^3},
\end{equation}
where $\epsilon_e$ is the first SR parameter at the end of the USR phase, i.e. $\epsilon_e = \big( \frac{\tau_e}{\tau_i} \big)^6 \epsilon_i$, $\tau_e$ is the conformal time at the end of the USR phase and $h$ is the relaxation parameter defined as,
\begin{equation}
\label{eq:h}
h = -6 \sqrt{\frac{\epsilon_V}{\epsilon_e}},
\end{equation}
with $\epsilon_V$ being the attractor of the second SR phase. According to the definition in Eq.~\eqref{eq:h}, we always have $h < 0$ and $|h|$ quantifies the speed of convergence to the second attractor; the larger $|h|$ is, the shorter the relaxation period will be.

Using once again the SR parameters associated with the USR phase Eq.~\eqref{eq:USR phase} and the subsequent SR regime, Eq.~\eqref{eq:Relaxation}, the SR parameters across the second transition can be expressed as

\begin{equation}
\text{For} \: \tau_e^- \leq \tau \leq \tau_e^+: \quad \epsilon = \epsilon_e, \quad \eta = -6 - h \theta(\tau - \tau_e).
\end{equation}

We are also interested in an special limit of this model where $\tau_e \to \tau_i$, in which the inflaton field transits instantly between two SR phases corresponding to the SR-SR model discussed in Sec.~\ref{sec:SR-SR}. The explicit form of the effective potential of this model can be found in Eq.~\eqref{eq:SR-SR Potential}. In this special limit, we have $\epsilon_i \to \epsilon_e$, and during the first and the second SR phases the evolution is the same as Eq.~\eqref{eq:SR-phase} and Eq.~\eqref{eq:Relaxation} respectively. The only difference is during the transition where,
\begin{equation}
   \label{eq:SR SR transition}
   \text{For}\: \tau_i^- \leq \tau \leq \tau_i^+ : \quad \epsilon = \epsilon_e , \quad \eta = -\left(6 +h\right) \theta(\tau - \tau_i)
\end{equation}

\subsection{Mode Functions}
\label{sec:mode functions}
Having found the full background solution, we now focus on the mode function of the curvature perturbation which satisfies the following equation of motion
\begin{equation}
   \label{eq:SM}
   \mathcal{R}^{\prime \prime}_k(\tau) - \frac{2+\eta}{\tau} \mathcal{R}^{\prime}_k(\tau) + k^2 \mathcal{R}_k (\tau) = 0,
\end{equation}
where prime denotes the derivative with respect to conformal time.
During the first SR phase,  the solution to the above equation takes the form  
\begin{equation}
\label{eq:R1}
   \mathcal{R}^{(1)}_k(\tau) = \frac{H}{\sqrt{4 \epsilon_i k^3}} \left( 1 + i k \tau \right) e^{-i k \tau}\, ,
\end{equation}
where the BD vacuum is assumed. 
%Note that the solution in Eq.~\eqref{eq:R1} is based on the assumption that $H$ is constant, while $\epsilon$ can vary arbitrarily.
During the USR and second SR phases, the inflationary vacuum is no longer the Bunch--Davies vacuum , since the inflaton filed has experienced sharp transitions. The mode function is instead given by  
\begin{equation}
   \mathcal{R}^{(2), (3)}_k = \frac{H}{\sqrt{4 \epsilon(\tau) k^3}} \left( \alpha^{(2), (3)}_k \left( 1 + i k \tau \right) e^{- i k \tau} + \beta^{(2), (3)}_k \left( 1- i k \tau \right) e^{i k \tau} \right)\, ,
\end{equation}
where $\alpha^{(2), (3)}_k$ and $\beta^{(2), (3)}_k$ are the Bogoliubov coefficients, labeled by indexes $(2)$ and $(3)$ for the second and third phases, respectively. These coefficients are obtained by requiring that both $\mathcal{R}$ and $\mathcal{R}^{\prime}$ remain continuous across the transitions. Explicitly, they are given by \cite{Firouzjahi:2025ihn},
\begin{equation}
\label{eq:alpha2 and beta2}
   \alpha^{(2)}_k = 1 + \frac{3 i}{2 k^3 \tau_i^3} \left(1 + k^2 \tau_i^2 \right), \quad \beta^{(2)}_k = \frac{- 3 i }{2 k^3 \tau_i^3} (1+ i k \tau_i)^2 e^{-2 i k \tau_i}\, ,
\end{equation}
and
\begin{equation}
\label{eq:alpha3 and beta3}
\begin{pmatrix}
\alpha^{(3)}_k \\
\\
\beta^{(3)}_k
\end{pmatrix}
=
\begin{pmatrix}
\gamma_{11} \quad \gamma_{12}\\
\\
\gamma_{21} \quad \gamma_{22}
\end{pmatrix}
\begin{pmatrix}
\alpha^{(2)}_k \\
\\
\beta^{(2)}_k
\end{pmatrix},
\end{equation}
where
\begin{equation}
\begin{aligned}
\gamma_{11} = \gamma_{22}^* = 1 + \frac{i h \left( 1 + k^2 \tau_e^2 \right)}{4 k^3 \tau_e^3}, \qquad 
\gamma_{12} = \gamma_{21}^* = \frac{i h e^{2 i k \tau_e}}{4 k^3 \tau_e^3} \left( 1 - i k \tau_e \right)^2.
\end{aligned}
\end{equation}
Furthermore, one can show that the power spectrum at the end of inflation takes the form  
\begin{equation}
\label{eq:power spectrum}
   \mathcal{P}_{k} = \frac{H^2}{8 \pi^2 \epsilon_V} |\alpha_k^{(3)} + \beta_k^{(3)}|^2.
\end{equation}

%It is worth noting that Eq.~\eqref{eq:power spectrum} can be generalized for any given model, by substituting the Bogoliubov coefficients 
%of the final phase of  inflation and the value of the first SR parameter at the end of inflation.
The mode function for the SR-SR model before the transition is given by Eq.~\eqref{eq:R1}. After the transition, the mode function is given by $\mathcal{R}^{(3)}$ after taking the limit $\tau_e \to \tau_i$.
% the Bogoliubov coefficients in the second SR phase reduce to  
%\begin{equation}
%   \alpha^{(2)}_k = 1 + \frac{i\left( 6 +h \right)}{4 k^3 \tau_i^3} \left(1 + k^2 \tau_i^2 \right) , \quad \beta_k^{(2)} = \frac{-i \left(6 + h \right) }{4 k^3 \tau_i^3} (1+ i k \tau_i)^2 e^{-2 i k \tau_i}.
%\end{equation}
%
%Once again, it should be emphasized that in this scenario as well, the power spectrum is determined by Eq.~\eqref{eq:power spectrum}, substituting the indexes $(3)$ with $(2)$.

\subsection{Cubic Action}
\label{sec:cubic action}
We recall that the general form of the cubic action can be written as \cite{Maldacena_2003, Chen_2010}
\begin{equation}
\label{eq:cubic action}
    \begin{aligned}
& S_3=\int d t d^3 x\left[a^3 \epsilon^2 \mathcal{R} \dot{\mathcal{R}}^2+a \epsilon^2 \mathcal{R}(\partial \mathcal{R})^2-2 a \epsilon \dot{\mathcal{R}}(\partial \mathcal{R})(\partial \chi)+\frac{a^3 \epsilon}{2} \dot{\eta} \mathcal{R}^2 \dot{\mathcal{R}} \right. \\
&+\frac{\epsilon}{2 a}(\partial \mathcal{R})(\partial \chi) \partial^2 \chi  \left.+\frac{\epsilon}{4 a}\left(\partial^2 \mathcal{R}\right)(\partial \chi)^2+\left.2 f(\mathcal{R}) \frac{\delta L}{\delta \mathcal{R}}\right|_1\right],
\end{aligned}
\end{equation}
%%%%%%%%%%%%%%%%%%%%%%%%%%%%%%%%%%%%%%%%%See README file, note 5
%\begin{equation}
%\label{eq:cubic action}
%S_3=\int d t d^3 x\left[\frac{a^3 \epsilon}{2} \dot{\eta} \mathcal{R}^2 \dot{\mathcal{R}}+2 f(\mathcal{R}) \left. \frac{\delta L}{\delta \mathcal{R}}\right|_1\right]+\mathcal{O}(\epsilon^2)
%\end{equation}
%%%%%%%%%%%%%%%%%%%%%%%%%%%%%%%%%%%%%%%%%%%%%%%%%%%%%%%%
where
\begin{equation}
\partial^2 \chi=a^2 \epsilon \dot{\mathcal{R}},\left.\quad \frac{\delta L}{\delta \mathcal{R}}\right|_1=a\left(\partial^2 \dot{\chi}+H \partial^2 \chi-\epsilon \partial^2 \mathcal{R}\right)\, ,
\end{equation}
and
\begin{equation}
\begin{aligned}
f(\mathcal{R}) & =\frac{\eta}{4} \mathcal{R}^2+\frac{1}{H} \mathcal{R} \dot{\mathcal{R}} +\frac{1}{4 a^2 H^2}\left[-(\partial \mathcal{R})(\partial \mathcal{R})+\partial^{-2}\left(\partial_i \partial_j\left(\partial_i \mathcal{R} \partial_j \mathcal{R}\right)\right)\right]  \\
& +\frac{1}{2 a^2 H}\left[(\partial \mathcal{R})(\partial \chi)-\partial^{-2}\left(\partial_i \partial_j\left(\partial_i \mathcal{R} \partial_j \chi\right)\right)\right] .
\end{aligned}
\end{equation}
Note that the last term in Eq.~\eqref{eq:cubic action} can be removed by a field redifinition, $\mathcal{R} \to \mathcal{R}_n + f(\mathcal{R}_n)$. This field redifinition can be carried out at the end of inflation, where all modes of interest are superhorizon and only the first two terms in $f(\mathcal{R})$ survive. From what remains in the cubic action, it is easy that only the term proportional to $\dot \eta$ can be relevant to all scenarios we consider in this paper. Therefore, we are left with two terms in the field redefinition and a cubic action of the form $S_3 \supset \int d \tau d^3 x \frac{a^2 \epsilon}{2} \eta^{\prime} \mathcal{R}^2 \mathcal{R}^{\prime}$. The former is the dominant contribution in the non-BD USR model while the latter dominates in the SR-SR and transient USR model.

%has $\eta' = 0$, and the leading contributions originate solely from the field redefinition terms. 
%As a result, the bispectrum in this case takes the simple form
%\begin{equation}
%   B_\mathcal{R}(k_1,k_2,k_3) = 12 \pi^4 \left( \frac{\mathcal{P}_\mathcal{R}(k_1) \mathcal{P}_\mathcal{R}(k_2)}{k_1^3 k_2^3} + \mathrm{perms.} \right),
%\end{equation}
%where together with Eq.~\eqref{eq:power spectrum}, leads to Eq.~\eqref{eq:NBD bispectrum}.

\subsection{Bispectrum}
\label{sec:bispectrum}

In the non-BD USR model, since at late times the modes are at superhorizon scales, we can use the relation ${\mathcal R}' =-\frac{3}{\tau} {\mathcal R}$. Then from the field redefinition it is easy to see that the bispectrum is given by Eq.~\eqref{eq:NBD bispectrum}.

For the transient USR and the SR-SR models, using the in–in formalism, we have
\begin{equation}
\label{eq:bispectrumA}
   B_{\mathcal{R}}\left(k_1, k_2, k_3\right)=-2\operatorname{Im} \mathcal{R}_{k_1}\left(0\right) \mathcal{R}_{k_2}\left(0\right) \mathcal{R}_{k_3}\left(0\right) \int_{-\infty}^{0} d \tau a^2 \epsilon \eta^{\prime}\left[\mathcal{R}_{k_1}^*(\tau) \mathcal{R}_{k_2}^*(\tau) \mathcal{R}_{k_3}^{* \prime}(\tau)+2 \text { perm. }\right]
\end{equation} 
The above time integral can be naturally decomposed into contributions from each interval of the background evolution. 
We first carry out this computation explicitly for the transient USR model, and then turn to the special limit corresponding to the SR–SR  model.

The first SR and the USR phase of the transient USR model $( \tau \leq \tau_i ) $, does not contribute to the bispectrum since in these phases $\eta'=0$. Thus, we are left with the contributions from two sharp transitions and the relation period to the second SR phase. Around the transition time, Eq.~\eqref{eq:bispectrumA} yields
\begin{equation}
\label{eq:transition integral}
\begin{aligned}
    \int_{\tau_* - \delta \tau}^{\tau_* + \delta \tau} d \tau a^2 \epsilon \eta^{\prime} &\left[\mathcal{R}_{k_1}^*(\tau) \mathcal{R}_{k_2}^*(\tau) \mathcal{R}_{k_3}^{* \prime}(\tau)+2 \text { perm. }\right] = \\
    &\left. \left( a^2 \epsilon \left[\mathcal{R}_{k_1}^*(\tau) \mathcal{R}_{k_2}^*(\tau) \mathcal{R}_{k_3}^{* \prime}(\tau)+2 \text { perm. }\right] \right) \right|_{\tau = \tau_*} \times \Delta \eta
\end{aligned}
\end{equation}
where $\Delta \eta \equiv \eta(\tau_* + \delta \tau) - \eta(\tau_* - \delta \tau)$ and $\tau_*$ denotes the transition conformal time and, for the second equality, we have used the fact that all  parameters but $\eta'$ are continuous during the transition.  Using Eq.~\eqref{eq:transition integral}, the bispectrum at the transitions takes the following form,
\begin{equation}
\label{eq:bispectrum at transitions}
B_\mathcal{R}^{T}(k_1 , k_2 , k_3) = \frac{-2 \Delta \eta \epsilon(\tau_*)}{H^2 \tau_*^2} \mathrm{Im} \left[ P_{k_1}(\tau_* , 0) P_{k_2}(\tau_* , 0)   \mathcal{R}_{k_3}(0) \mathcal{R}_{k_3}^{*\prime}(\tau_*) + 2\mathrm{perms.}  \right].
\end{equation}
Accordingly, the contribution of the transition to the bispectrum will be as presented in Eq.~\eqref{eq:transient USR bT1} and Eq.~\eqref{eq:transient USR bT2}.

To perform the integral of the Eq.~\eqref{eq:bispectrumA} for the relaxation period, it is helpful to notice that the combination $a^2 \epsilon(\tau) \eta'(\tau) = \frac{h (h+6) }{2 H^2 \tau_e^3}$ is a constant throughout the relaxation phase. Thus,
\begin{equation}
\label{eq:relaxation integral}
 \int_{\tau_e}^{0} d \tau a^2 \epsilon \eta^{\prime} \left[\mathcal{R}_{k_1}^*(\tau) \mathcal{R}_{k_2}^*(\tau) \mathcal{R}_{k_3}^{* \prime}(\tau)+2 \text { perm. }\right]  = \frac{h (h+6) }{2 H^2 \tau_e^3} \left. \left( \mathcal{R}_{k_1}^*(\tau) \mathcal{R}_{k_2}^*(\tau) \mathcal{R}_{k_3}^*(\tau) \right) \right|_{\tau = \tau_e}^{\tau = 0},
\end{equation}
and hence the bispectrum will be,
\begin{equation}
\label{eq:bispectrum relaxation}
B_\mathcal{R}^{R}(k_1, k_2, k_3) =  \frac{h (h+6) }{ H^2 \tau_e^3} \mathrm{Im}\left[ P_{k_1}(\tau_e , 0) P_{k_2}(\tau_e , 0) P_{k_3}(\tau_e , 0) \right],
\end{equation}
where we have used the fact that the equal time power spectrum is real. 

The bispectrum for the SR-SR model  has two contributions, one from the transition and the other from the relaxation. The form of these contributions are the same as the ones presented in Eq.~\eqref{eq:bispectrum at transitions} and Eq.~\eqref{eq:bispectrum relaxation}. The final bispectrum of this model has been presented in Eq.~\eqref{eq:SR SR bispectrum}. This result may be reproduced by taking the limit of $\tau_e \to \tau_i$ and $\epsilon_e \to \epsilon_i$ in the full bispectrum for the transient USR model, presented in Eq.~\eqref{eq:transient USR bispectrum}.

%\section{Full Shape Correlator}
%\label{sec:full shape correlator}
%In Sec.~\ref{sec:Tools}, We have defined the shape correlator for each overall scale of the triangle, integrating over the relative configurations. However, it is useful to consider a full inner product which is marginalized over the total scale of the triangle as well. 
%This provides a global measure of how close a given model is to a specific template across all momentum scales.
%
%We define the full-scale inner product between two shapes $S_1$ and $S_2$ as,
%\begin{equation}
%\label{eq:full inner product}
%\langle S_1 , S_2 \rangle = \int_0^\infty \langle S_1 , S_2 \rangle_k \mathcal{W}_k \mathrm{d}k,
%\end{equation}
%where $\mathcal{W}_k$ is a suitable weight function.

\bibliographystyle{JHEP}
\bibliography{biblio.bib}

\end{document}